\documentclass{article}

\usepackage{PRIMEarxiv}
\usepackage[utf8]{inputenc}
\usepackage[T2A]{fontenc}
\usepackage[russian,english]{babel}
\usepackage[square, comma, numbers, sort&compress]{natbib}

\usepackage{hyperref}       % hyperlinks
\usepackage{url}            % simple URL typesetting
\usepackage{booktabs}       % professional-quality tables
\usepackage{amsfonts}       % blackboard math symbols
\usepackage{nicefrac}       % compact symbols for 1/2, etc.
\usepackage{microtype}      % microtypography
\usepackage{lipsum}
\usepackage{fancyhdr}       % header
\usepackage{graphicx}       % graphics
\usepackage{amsmath}  % For using math symbols
\usepackage{siunitx}  % For using SI units
\usepackage{multirow} % For merging rows in tables
\graphicspath{{media/}}     % organize your images and other figures under media/ folder
\usepackage[table]{xcolor}  % For adding color to tables
\usepackage{array}
\newcolumntype{C}[1]{>{\centering\arraybackslash}m{#1}}
\title{POINT$^{2}$: A Polymer Informatics Training and Testing Database
}

\author{
  Jiaxin Xu, Gang Liu, Ruilan Guo, Meng Jiang, Tengfei Luo* \\
  University of Notre Dame \\
  Notre Dame, IN, USA\\
  \texttt{*tluo@nd.edu} \\
}

\begin{document}
\maketitle

\begin{abstract}
The advancement of polymer informatics has been significantly propelled by the integration of machine learning (ML) techniques, enabling the rapid prediction of polymer properties and expediting the discovery of high-performance polymeric materials. However, the field lacks a standardized workflow that encompasses prediction accuracy, uncertainty quantification, ML interpretability, and polymer synthesizability. In this study, we introduce POINT$^{2}$ (POlymer INformatics Training and Testing), a comprehensive benchmark database and protocol designed to address these critical challenges. Leveraging the existing labeled datasets and the unlabeled PI1M dataset—a collection of approximately one million virtual polymers generated via a recurrent neural network trained on the realistic polymers—we develop an ensemble of ML models, including Quantile Random Forests, Multilayer Perceptrons with dropout, Graph Neural Networks, and pretrained large language models. These models are coupled with diverse polymer representations such as Morgan, MACCS, RDKit, Topological, Atom Pair fingerprints, and graph-based descriptors to achieve property predictions,  uncertainty estimations, model interpretability, and template-based polymerization synthesizability across a spectrum of properties, including gas permeability, thermal conductivity, glass transition temperature, melting temperature, fractional free volume, and density. The POINT$^{2}$ database can serve as a valuable resource for the polymer informatics community for polymer discovery and optimization.
\end{abstract}

% keywords can be removed

\section{Introduction}
\begin{figure}
    \centering
    \includegraphics[width=0.7\linewidth]{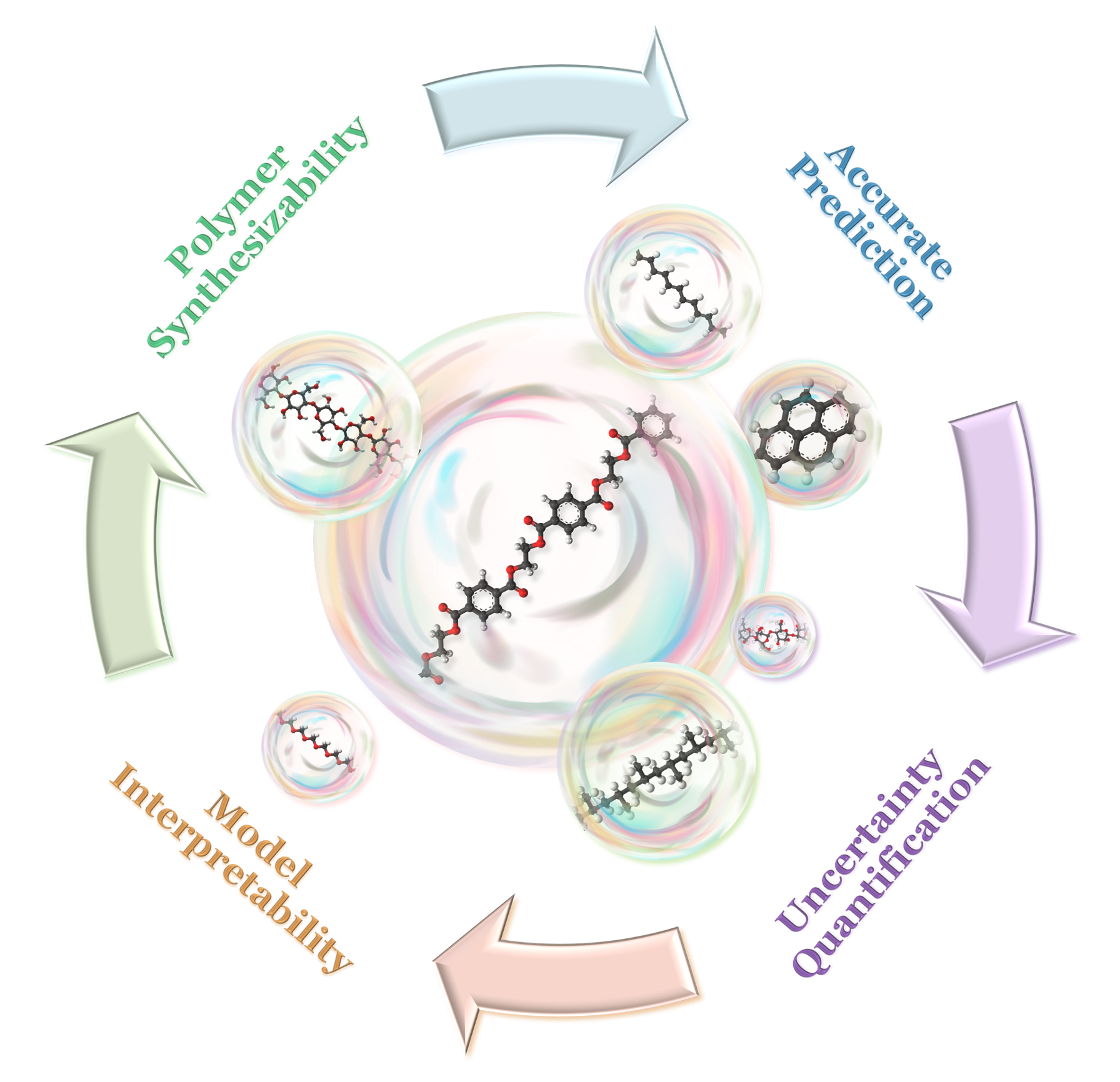}
    % \includegraphics[width=0.7\linewidth]{fig1_new.pdf}
    % \vspace{-0.1in}
    \caption{Schematic representation of the integrated workflow for polymer screening, highlighting four key components: Accurate Prediction, Uncertainty Quantification, Model Interpretability, and Polymer Synthesizability.}
    \label{fig:1}
    % \vspace{-0.25in}
\end{figure}

Polymers are essential to numerous industries such as renewable energy, biomedical devices, aerospace engineering, food, consumer goods, and advanced electronics, due to their diverse and tunable properties \cite{feldman2008polymer, koltzenburg2023polymer,mohanty2022sustainable}. The traditional trial-and-error approach to polymer development is often time-consuming and resource-intensive \cite{allcock1992rational,maranas1996optimal,gani2019group}. Recent advancements in polymer informatics have demonstrated the potential of data-driven methodologies, particularly machine learning (ML), to predict polymer properties and streamline the discovery of novel polymeric materials \cite{chen2021polymer,hatakeyama2023recent,tran2024design,xu2024transcend}. Despite significant advancements, polymer informatics faces critical challenges hindering its progress towards robust, automated material discovery. Key among these is the absence of standardized workflows that effectively integrate prediction accuracy, uncertainty quantification (UQ), model interpretability, and polymer synthesizability—four elements that we propose to address in this work, as shown in Fig. \ref{fig:1}.

\subsection{Prediction Accuracy}
The capability to accurately predict polymer properties using ML models is a cornerstone of polymer informatics. While numerous studies showcase the application of various ML algorithms for this purpose \cite{ma2022machine, liu2022graph, doan2020machine, xu2024superior, aldeghi2022graph, xu2024unlocking, park2022prediction,tao2023machine, yang2022machine,yamada2019predicting, wu2019machine}, the field lacks standardized benchmark datasets. Unlike in areas such as small molecule discovery \cite{wu2018moleculenet, polykovskiy2020molecular, gaulton2012chembl, gaulton2017chembl}, natural language processing \cite{wang2018glue, wang2019superglue,rajpurkar2016squad, bowman2015large}, and computer vision \cite{deng2009imagenet,lin2014microsoft,krizhevsky2009learning}, where well-designed benchmarking datasets facilitate fair comparisons of model performance, polymer informatics has not yet established similar resources. Moreover, large language models (LLMs) have demonstrated remarkable capabilities across diverse domains \cite{zhao2023survey, chang2024survey, thirunavukarasu2023large}. In chemistry, they have shown promise in tasks such as molecular property prediction \cite{castro2023large, jablonka2024leveraging, guo2023can}, retrosynthetic analysis \cite{m2024augmenting}, and inverse design \cite{jablonka2024leveraging}. However, their application to polymer property prediction remains largely unexplored, necessitating further investigation into their capacity to learn and generalize polymer-specific structure-property relationships. The creation of standardized benchmarks is crucial as they would not only validate the efficacy of predictive models but also promote transparency and reproducibility in research. Additionally, such benchmarks would provide a foundation for systematically evaluating the potential of emerging technologies, e.g., LLMs, in polymer informatics. There is a clear need for a comprehensive database to further fuel innovation in the field of polymer informatics.

\subsection{Uncertainty Quantification}
Current research in polymer informatics primarily focuses on predictive accuracy, often overlooking the critical aspect of UQ. UQ sheds light on the confidence of model predictions by addressing both \textit{aleatoric uncertainty}, which arises from inherent variability in polymeric materials and measurement processes, and \textit{epistemic uncertainty}, stemming from model limitations and incomplete knowledge of the input space \cite{soize2017uncertainty, abdar2021review, sullivan2015introduction}. In polymer science, aleatoric uncertainty can emerge from multiple stages of the materials pipeline, including monomer variability, polymerization pathways, molecular weight distributions, and processing and measurement conditions, all of which influence final properties such as thermal conductivity, glass transition temperature, and gas permeability \cite{jha2019impact, patrone2016uncertainty, tang2024uncertainty}. However, this work focuses specifically on epistemic uncertainty, as the datasets available do not contain sufficient information to model aleatoric uncertainty—such as replicate measurements or labeled noise distributions. In ML, epistemic uncertainty is commonly estimated through Bayesian methods like Gaussian Processes \cite{rasmussen2003gaussian,mackay1998introduction} and Bayesian Neural Networks \cite{jospin2022hands,goan2020bayesian,izmailov2021bayesian}, which capture parameter uncertainty directly, or through ensemble methods such as Random Forests \cite{breiman2001random,mentch2016quantifying} and Monte Carlo Dropout \cite{gal2016dropout,gal2017concrete}, which assess uncertainty by analyzing variability across multiple predictions. 

Beyond enhancing model confidence, UQ also serves as a valuable feedback mechanism in iterative model development and decision-making frameworks, including active learning \cite{settles2009active,settles2011theories}, Bayesian optimization \cite{pelikan2005bayesian, shahriari2015taking}, and reinforcement learning \cite{wiering2012reinforcement,arulkumaran2017deep}. This is especially important in polymer discovery, where models must often extrapolate into sparsely sampled or out-of-distribution regions to identify novel, high-performance polymers. In such scenarios, reliable uncertainty estimates are essential for evaluating model trustworthiness and guiding experimental validation. Ultimately, integrating UQ into automated polymer property prediction workflows is essential—not only to improve robustness and transparency, but also to support risk-aware decisions in materials design. UQ should be treated as a first-class metric, on par with accuracy, in advancing the next generation of polymer informatics.

\subsection{Model Interpretability}
As ML models grow in complexity, they often become more opaque, earning the label of "black boxes." However, interpretability is crucial for gaining user trust and enabling developers to effectively monitor and refine these models \cite{carvalho2019machine,gilpin2018explaining}. It also plays a key role in deciphering the complex mechanisms underlying big data, potentially leading to new scientific discoveries \cite{dybowski2020interpretable, makke2024symbolic, 9007737}. Understanding how models make their predictions allows researchers to gain deeper insights into polymer behaviors and the intricate relationships between their structures and properties \cite{babbar2024explainability,tao2023discovery,yang2022machine,xu2024superior,park2022prediction}. This advancement enriches both the scientific understanding and practical applications of polymer informatics. There are two main types of interpretability methods for ML models: \textit{model-agnostic} and \textit{model-intrinsic}. Model-agnostic methods, such as LIME (Local Interpretable Model-agnostic Explanations) \cite{ribeiro2016should} and SHAP (SHapley Additive exPlanations) \cite{scott2017unified}, provide insights regardless of the model architecture. Model-intrinsic methods, like linear models, tree-based models \cite{lundberg2020local,costa2023recent}, attention mechanisms \cite{vaswani2017attention,niu2021review,bahdanau2014neural,xu2015show,wiegreffe2019attention}, and rationalization \cite{gurrapu2023rationalization,wong2024discovery,liu2022graph,wu2022discovering}, offer interpretability directly embedded within the model's structure, facilitating a more integrated understanding of prediction processes.

\subsection{Synthesizability}
Finally, even if the aforementioned challenges are addressed, the practical application of predicted polymer structures remains contingent upon their synthesizability. The integration of synthesizability assessments, especially polymerization assessments, into the workflow ensures that the predicted polymers are not only theoretically optimal but also practically synthesizable. Unlike small molecules, polymers require specific polymerization steps that must be feasible. Polymerization is the chemical process in which small molecules, known as monomers, combine to form larger, chain-like or network structures of polymers \cite{odian2004principles,carothers1931polymerization}. In small organic molecule design, various methods have been developed to predict or evaluate synthesizability. Retrosynthesis planning, encompassing both template-based and template-free approaches, deconstructs a target molecule into simpler precursor structures, effectively mapping a synthetic route \cite{corey1991logic,corey1988robert,sun2022computer,liu2024multimodal, chen2024reaction}. Template-based methods utilize predefined or retrieved reaction templates to guide the deconstruction process \cite{baylon2019enhancing,segler2017neural,coley2017computer,dai2019retrosynthesis,chen2021deep}, while template-free methods employ ML techniques, e.g., deep generative models, to generate retrosynthetic pathways without relying on explicit templates \cite{lin2020automatic,liu2017retrosynthetic,zheng2019predicting,tetko2020state,zhong2022root,tu2022permutation,wan2022retroformer,liu2024mars,zhong2023retrosynthesis}. Another approach to evaluating synthesizability, without proposing explicit reaction routes, involves structural complexity-based scoring systems. Methods such as the Synthetic Accessibility Score (SAScore) \cite{ertl2009estimation} assess the ease of synthesis based on molecular complexity and the presence of challenging substructures. Variants of SAScore include the SCScore \cite{coley2018scscore}, Synthetic Bayesian Accessibility (SYBA) \cite{vorvsilak2020syba}, and Graph Attention-based assessment of Synthetic Accessibility (GASA) \cite{yu2022organic}. These methodologies collectively contribute to the understanding of the synthesizability of small organic molecules, bridging theoretical predictions and practical applications. 

However, these approaches often do not extend easily to polymers due to the complexities inherent in polymerization processes, which are not included in the synthesizability assessments of small organic molecules. Generative models, such as deep generative models, have shown promise in designing novel polymers with desired properties \cite{chen2020machine,sattari2021data,batra2020polymers,ma2020pi1m,kim2023open,gurnani2021polyg2g,qiu2024demand,liu2023high,huang2024ai,yang2024novo}. Nevertheless, without incorporating polymer-specific synthesizability assessments, these models may propose structures that are challenging or impractical to synthesize. To the best of our knowledge, the work by Chen et al. \cite{chen2021data} is the only study specifically focused on polymer retrosynthesis planning, taking into account the unique complexities of polymerization processes. In their study, Chen et al. manually compiled a comprehensive dataset of polymerization reactions from various resources, extracting hundreds of synthetic templates that interpret the chemical reactions between reactant monomers. Utilizing this dataset, they built a polymer retrosynthesis framework that employs a similarity metric to select synthetic pathways for target polymers, facilitating the prediction of feasible synthesis routes. However, the retrosynthesis tool and the underlying polymer reaction data are not open-sourced and can only be accessed via a web interface \cite{doan2020machine}, limiting exhaustive searches and broader accessibility. Additionally, the web interface may encounter challenges with complex polymer structures, leading to extended computation times and instability.

\medskip % This adds a smaller vertical space

In this study, we present the POINT$^{2}$ (POlymer INformatics Training and Testing) framework, a benchmark database and protocol designed for polymer property prediction and screening. We utilize an ensemble of ML models, including Quantile Random Forests (QRF) \cite{meinshausen2006quantile}, Multilayer Perceptrons with dropout (MLP-D) \cite{gal2016dropout}, and various Graph Neural Networks (GNN), like Graph Isomorphism Networks (GIN) \cite{xu2018powerful}, Graph Convolutional Networks (GCN) \cite{kipf2016semi}, and Graph Rationalization with Environment-based Augmentations (GREA) \cite{liu2022graph}), chosen for their robust accuracy and capacity for uncertainty estimation.In addition to these ML models, we also evaluate the predictive capabilities of pretrained LLMs in an in-context learning (ICL) setting~\cite{dong2022survey}, comparing their performance against traditional ML approaches to assess the potential of LLMs in polymer informatics and property prediction. These models are integrated with interpretable polymer representations such as Morgan, MACCS, RDKit, Topological, Atom Pair fingerprints, and graph-based descriptors, ensuring accurate and interpretable results. Our predictive tasks target key polymer properties such as gas permeability (P) for five major industrial gases (O$_2$, N$_2$, O$_2$, CH$_2$, H$_2$, and CO$_2$), thermal conductivity (TC), glass transition temperature (T\textsubscript{g}), melting temperature (T\textsubscript{m}), fractional free volume (FFV), and density (\( \rho \)), chosen for their relatively extensive data availability and their significant roles in practical applications across various industries. Alongside these capabilities, we introduce an open-sourced, template-based retrosynthesis tool for polymers, equipped with a synthesizability score tailored for polymers (PolyScore), which can aid chemists in synthetic planning and evaluate the results of polymer generative models. Our models systematically screen the PI1M \cite{ma2020pi1m} database—a collection of approximately one million hypothetical polymers generated using a recurrent neural network (RNN) trained on the realistic polymers—delivering comprehensive results that include predicted properties, their uncertainties, model prediction interpretation, and polymer synthesizability. Additionally, the curated labeled dataset provided by POINT$^{2}$ can serve as a benchmark resource for future work in polymer informatics and the evaluation of ML algorithms, establishing a foundational tool for the community’s ongoing research and development efforts.
% \section{Headings: first level}
% \label{sec:headings}

% % See Section \ref{sec:headings}.

% \subsection{Headings: second level}

% % \begin{equation}
% % \xi _{ij}(t)=P(x_{t}=i,x_{t+1}=j|y,v,w;\theta)= {\frac {\alpha _{i}(t)a^{w_t}_{ij}\beta _{j}(t+1)b^{v_{t+1}}_{j}(y_{t+1})}{\sum _{i=1}^{N} \sum _{j=1}^{N} \alpha _{i}(t)a^{w_t}_{ij}\beta _{j}(t+1)b^{v_{t+1}}_{j}(y_{t+1})}}
% % \end{equation}

% \subsubsection{Headings: third level}

% \paragraph{Paragraph}

\section{Results and Discussion}
\label{sec:Results}
\subsection{Benchmark Dataset}

This study utilizes a comprehensive dataset encompassing a diverse range of polymer properties essential for different applications, such as renewable energy, biomedical devices, aerospace engineering, and advanced electronics. As summarized in Table \ref{tab:dataset}, the benchmark dataset includes properties obtained through both experimental measurements and computational simulations. Specifically, T\textsubscript{g}, T\textsubscript{m}, $\rho$, and P (for O$_2$, O$_2$, CH$_4$, H$_2$, and CO$_2$ gases) are experimentally measured labels, collected from established databases like MSA \cite{thornton2012polymer}, along with additional literature sources \cite{xu2024superior, tao2021benchmarking, uddin2024interpretable,tao2021machine, jha2019impact,kim2018polymer,bicerano2002prediction,wypych2022handbook,van2009properties,mark2009polymer}. FFV and TC were derived from molecular dynamics (MD) simulations in previous studies \cite{tao2023machine, ma2022machine}. The data for each property were cleaned and randomly split into training and testing sets in a 4:1 ratio. Further chemical space and property space distribution comparison of the training and testing data for each property can be found in Figs. \ref{fig:tsne} and \ref{fig:label_dist}.

Specifically, the polymer properties featured in this benchmark dataset represent a hierarchical level of difficulty in prediction. $\rho$, for instance, is closely related to more intrinsic aspects of the polymer's chemical structure \cite{van1969prediction}, whereas T\textsubscript{g} is influenced by the polymer's compositional and configurational characteristics \cite{babbar2024explainability}. TC presents a greater challenge, as it involves complex interactions at the molecular level, which are often more difficult to predict accurately compared to bulk properties like density \cite{ma2022machine}. P, involving mass transport phenomena through polymer matrices, is influenced by both the molecular structure and the interaction of gases with the polymer, making it among the most complex properties to predict \cite{sanders2013energy}.

We expect this high-quality and diverse dataset to serve as a valuable benchmark for the polymer informatics community and beyond, enabling researchers to train models and evaluate performance on a standardized split, thereby promoting fair and consistent model comparisons.

% \begin{table}[ht]
% \centering
% \caption{Dataset Statistics}
% \label{tab:dataset}
% \begin{tabular}{|l|c|c|c|}
% \toprule
% \textbf{Properties} & \textbf{Units} & \textbf{\# Training} & \textbf{\# Testing} \\
% \hline
% \hline
% T\textsubscript{g} & \si{\celsius} & 5766 & 1442 \\
% \hline
% T\textsubscript{m} & \si{\celsius} & 2936 & 735 \\
% \hline
% TC & \si{W/(m.K)} & 1264 & 316 \\
% \hline
% FFV & 1 & 6436 & 1610 \\
% \hline
% $\rho$ & \si{g/cm^3} & {---} & {---} \\
% \hline
% P$_{O_2}$ & \multirow{5}{*}{$\log_{10}(\text{Barrer})$} & 644 & 161 \\
% \cline{1-1} \cline{3-4}
% P$_{N_2}$ & & 635 & 159 \\
% \cline{1-1} \cline{3-4}
% P$_{CH_4}$ & & 544 & 137 \\
% \cline{1-1} \cline{3-4}
% P$_{H_2}$ & & 407 & 102 \\
% \cline{1-1} \cline{3-4}
% P$_{CO_2}$ & & 603 & 151 \\
% \bottomrule
% \end{tabular}
% \end{table}

\begin{table}[ht]
\centering
\caption{Summary of Properties in the Benchmark Polymer Dataset. The dataset includes glass transition temperature (T\textsubscript{g}), melting temperature (T\textsubscript{m}), thermal conductivity (TC), fractional free volume (FFV), density ($\rho$), and gas permeabilities (P) for O$_2$, N$_2$, CH$_4$, H$_2$, and CO$_2$, with data split into training and testing sets at a 4:1 ratio.}
\label{tab:dataset}
\begin{tabular}{@{} l c r r @{}}
\toprule
\textbf{Properties} & \textbf{Units} & \textbf{\# Training} & \textbf{\# Testing} \\
\midrule
\rowcolor{gray!10}
T\textsubscript{g}          & \si{\celsius}              & 5,766 & 1,442 \\
T\textsubscript{m}          & \si{\celsius}              & 2,936 & 735  \\
\rowcolor{gray!10}
TC                          & \si{W/(m.K)}               & 1,264 & 316  \\
FFV                         & 1                          & 6,436 & 1,610 \\
\rowcolor{gray!10}
$\rho$                      & \si{g/cm^3}                & 1,368 & 342 \\
\hline
P(O$_2$)                   & \multirow{5}{*}{$\log_{10}(\text{Permeability in Barrer})$} & 644  & 161  \\
\rowcolor{gray!10}
P(N$_2$)                 &                             & 635  & 159  \\
P(CH$_4$)                 &                             & 544  & 137  \\
\rowcolor{gray!10}
P(H$_2$)                  &                             & 407  & 102  \\
P(CO$_2$)                &                             & 603  & 151  \\

\bottomrule
\end{tabular}
\end{table}

\subsection{Prediction Accuracy Comparison}
We trained multiple ML models, including QRF, MLP-D, vanilla GNN (GIN and GCN), and augmented GNN (GREA), using the training sets. QRF and MLP-D were evaluated with various polymer fingerprinting methods (Morgan, MACCS, RDKit, Topological, and Atom Pair). Graph-based models (GIN, GCN, and GREA) were directly trained on the graph description of polymer structures. The performance of these models was then assessed on the hold-out test sets. To further investigate alternative predictive approaches, we evaluated the pretrained large language model (LLM), GPT-4o-mini~\cite{hurst2024gpt}, under zero-shot and few-shot in-context learning (ICL) settings, with 0, 5, 10, and 20 ICL examples (randomly sampled from the training set) provided per test sample. Detailed methodologies for polymer representation, model training, and evaluation are provided in Sections \ref{sec:representation} and \ref{sec:MLmodel}.

\begin{table}[h]
\centering
\caption{Comparison of Model Prediction Performance on Testing Dataset. All values reported are Root Mean Square Errors (RMSE) in units corresponding to each property. The top two models for each property are underscored. `avg' denotes the mean RMSE for each model across all fingerprinting methods. `TT' and `AP' are short for Topological Torsion fingerprint and Atom Pair fingerprint, respectively. Numbers (0, 5, 10, and 20) after `GPT-4o-mini' denotes the number of examples used in ICL. `Training (avg)' refers to the RMSE obtained when using the training set mean as the prediction for the test set.}

\label{tab:test_acc_comparison}
% \begin{tabular}{@{}lSSSSSSSSSS@{}}
\begin{tabular}{
  @{} 
  l 
  S[table-format=2.2] 
  S[table-format=2.2] 
  S[table-format=1.3] 
  S[table-format=1.3] 
  c 
  S[table-format=1.3] 
  S[table-format=1.3] 
  S[table-format=1.3] 
  S[table-format=1.3] 
  S[table-format=1.3] 
  @{}
}
\toprule
\textbf{Models} & {T\textsubscript{g}} & {T\textsubscript{m}} & {TC} & {FFV} & {$\rho$} & {P(O$_2$)} & {P(N$_2$)} & {P(H$_2$)} & {P(CH$_4$)} & {P(CO$_2$)} \\
\midrule
QRF-Morgan       & 38.07 & 59.79 & 0.056 & 0.016 & 0.107 & 0.698 & 0.577 & 0.436 & 0.606 & 0.690 \\
QRF-MACCS        & 43.53 & 61.15 & 0.055 & 0.016 &  \underline{0.101} & 0.760 & 0.627 & \underline{0.429} & 0.809 & 0.723 \\
QRF-RDKit        & 39.23 & 60.13 & 0.056 & 0.016 & 0.118 & 0.716 & 0.533 & 0.509 & 0.593 & 0.814 \\
QRF-TT           & 39.27 & 62.48 & 0.051 & 0.017 & 0.182 & 0.805 & 0.526 & 0.517 & 0.550 & 0.875 \\
QRF-AP           & 40.05 & 61.69 & 0.049 & 0.016 & 0.120 & 0.726 & 0.505 & 0.514 & 0.568 & 0.774 \\
\hline
\rowcolor{gray!20} % Adds a light gray background to the QRF (avg) row
QRF (avg)          & 40.03 & 61.05 & 0.053 & 0.016 & 0.126 & 0.741 & 0.554 & 0.481 & 0.625 & 0.775 \\
\hline
MLP-D-Morgan     & 37.57 & 59.47 & \underline{0.047} & 0.015 & 0.117 & 0.640 & \underline{0.490} & 0.478 & 0.590 & 0.646 \\
MLP-D-MACCS      & 40.42 & 61.06 & 0.059 & \underline{0.013} &  \underline{0.104} & 0.695 & 0.601 & \underline{0.371} & 0.702 & 0.704 \\
MLP-D-RDKit      & 38.18 & 58.87 & 0.057 & \underline{0.014} & 0.126 & 0.739 & 0.545 & 0.469 & 0.606 & 0.742 \\
MLP-D-TT         & 39.37 & 63.16 & \underline{0.047} & 0.016 & 0.126 & 0.689 & 0.534 & 0.486 & 0.595 & 0.728 \\
MLP-D-AP         & 38.55 & 59.41 & 0.052 & \underline{0.014} & 0.111 & 0.631 & 0.524 & 0.437 & 0.633 & \underline{0.628} \\
\hline
\rowcolor{gray!20} % Adds a light gray background to the QRF (avg) row
MLP-D (avg)        & 38.82 & 60.39 & 0.052 & 0.014 & 0.117 & 0.679 & 0.539 & 0.448 & 0.625 & 0.690 \\
\hline
GNN              & \underline{36.01} & \underline{55.47} & 0.077 & 0.021 & 0.168 & \underline{0.608} & \underline{0.486} & 0.469 & \underline{0.468} & \underline{0.618} \\
GREA             & \underline{37.32} & \underline{57.10} & 0.066 & 0.023 & 0.126 & \underline{0.566} & 0.513 & 0.447 & \underline{0.549} & 0.634 \\
\hline
GPT-4o-mini-0    & 100.92 & 110.47 & 0.112 & 0.178 & 0.189 & 2.949 & 3.023 & 4.684 & 3.212 & 2.033 \\
GPT-4o-mini-5    & 95.54 & 114.75 & 0.096 & 0.039 & 0.182 & 1.440 & 1.629 & 1.198 & 1.656 & 1.327 \\
GPT-4o-mini-10   & 91.11 & 111.03 & 0.092 & 0.035 & 0.172 & 1.320 & 1.520 & 1.170 & 1.533 & 1.290 \\
GPT-4o-mini-20   & 85.98 & 105.65 & 0.083 & 0.031 & 0.169 & 1.267 & 1.381 & 1.064 & 1.342 & 1.257 \\
\hline
\rowcolor{gray!20} % Adds a light gray background to the QRF (avg) row
Training (avg)        & 111.57 & 113.00 & 0.089 & 0.030 & 0.194 & 1.323 & 1.430 & 1.167 & 1.426 & 1.285 \\
\bottomrule
\end{tabular}
\end{table}

The model performance on the test sets, summarized in Table \ref{tab:test_acc_comparison}, indicates that graph-based models (vanilla GNN and GREA) generally outperformed other models across most tasks (6 out of 10). This superior performance can be attributed to GNNs' ability to effectively capture the intricate topological and relational information inherent in polymer molecules, enabling a more natural capture of the atomic connectivity that dictates different properties. Comparatively, MLP-D models demonstrated better predictive accuracy than QRF models in almost all tasks. This may stem from MLP-D's capacity to model complex, non-linear relationships within the data, facilitated by its deep learning architecture and the incorporation of dropout for regularization, which enhances generalization to unseen data.

Regarding the pre-trained LLM approach, it showed some predictive capability relative to the simplest baseline—using the training set mean as the prediction for test data—particularly for properties such as T\textsubscript{g}, T\textsubscript{m}, and $\rho$. Notably, its zero-shot performance on these three properties surpassed the training average baseline, which may be attributed to the hierarchical level of prediction difficulty discussed in the previous section, where these properties were found to be relatively easier to predict. Moreover, increasing the number of ICL examples (from 0 to 20) consistently improved predictive performance across all properties. For instance, RMSE values dropped significantly from the 0-shot to the 20-shot setting, suggesting that pre-trained LLMs benefit substantially from additional contextual information. Except for FFV, all tasks achieved better performance with 20-shot ICL than with the training average method, despite the latter leveraging the full training dataset (as shown in Table \ref{tab:dataset}). However, despite these improvements, LLMs-ICL were still less competitive than dedicated ML models, which were explicitly trained on large polymer datasets.

In evaluating the average performance of various polymer fingerprints, as shown in Appendix Table \ref{tab:fingerprint_averages}, the frequency with which each fingerprint achieved the top rank is as follows: Morgan (T\textsubscript{g}, P(O$_2$), and P(CO$_2$)) = MACCS (FFV, $\rho$, and P(H$_2$)) > Topological Torsion (TC and P(CH$_4$)) > RDKit (T\textsubscript{m}) = Atom Pair (P(N$_2$)). The comparable performance across different fingerprints indicates that no single representation universally outperforms the others across all tasks. Each fingerprint's design emphasizes different aspects of molecular structure, making them more or less suitable for predicting specific properties. Therefore, selecting the most appropriate fingerprint may depend on the particular property of interest and the underlying structural features that influence it. The efficacy of different fingerprints appears to be influenced by the hierarchical complexity of the properties being predicted. For properties like  $\rho$, FFV, and T\textsubscript{g}, which are closely linked to the polymer's intrinsic chemical structure, fingerprints that effectively capture local chemical environments, such as Morgan and MACCS, tend to yield better results. On the other hand, properties like TC and P, which often involve complex long-range interactions and depend on both local and global structural features, may require representations that strike a balance between capturing detailed local information and broader topological features, like Morgan, Topological Torsion, and Atom Pair. This underscores the need for more advanced fingerprints or hybrid representations that can encapsulate both fine-grained substructural details and higher-order molecular topology for accurate prediction of challenging polymer properties.

\subsection{Prediction Uncertainty Quantification}

Evaluating UQ is crucial for assessing the confidence of model predictions, especially when selecting candidate polymers for experimental validation in subsequent steps. Intuitively, we expect that the predicted uncertainty should provide meaningful coverage of the prediction error, i.e., the difference between the predicted value and the ground truth. This implies that a reasonable uncertainty estimate should increase in regions where the model is likely to make larger errors. In this case, UQ acts as a measure of the model’s reliability—when uncertainty is high, the model's prediction is less trustworthy, whereas low uncertainty indicates higher confidence in the predicted value. We employed two metrics to evaluate the quality of UQ: Spearman's rank correlation coefficient and sparsification plots.

\begin{table}[h]
\centering
\caption{Comparison of Model Uncertainty Quantification on Testing Dataset. All values reported are the Spearman's rank correlation coefficient ($\rho_s$) between prediction error and prediction uncertainty. The top two models with the highest $\rho_s$ for each property are underscored. `avg' denotes the mean coefficient value for each model across all fingerprinting methods. `TT' and `AP' are short for Topological Torsion fingerprint and Atom Pair fingerprint, respectively.}
\label{tab:test_uncertain}
% \begin{tabular}{@{}lSSSSSSSSSS@{}}
\begin{tabular}{
  @{} 
  l 
  S[table-format=1.3] 
  S[table-format=1.3] 
  S[table-format=1.3] 
  S[table-format=1.3] 
  c 
  S[table-format=1.3] 
  S[table-format=1.3] 
  S[table-format=1.3] 
  S[table-format=1.3] 
  S[table-format=1.3] 
  @{}
}
\toprule
\textbf{Models} & {T\textsubscript{g}} & {T\textsubscript{m}} & {TC} & {FFV} & {$\rho$} & {P(O$_2$)} & {P(N$_2$)} & {P(H$_2$)} & {P(CH$_4$)} & {P(CO$_2$)} \\
\midrule
QRF-Morgan       & 0.317 & 0.382 & \underline{0.338} & \underline{0.385} & 0.380 & \underline{0.461} & \underline{0.386} & 0.376 & \underline{0.291} & \underline{0.455} \\
QRF-MACCS        & 0.283 & 0.314 & 0.246 & 0.355 & 0.364 & 0.272 & 0.361 & 0.254 & 0.272 & 0.330 \\
QRF-RDKit        & 0.372 & 0.387 & 0.323 & 0.377 & 0.370 & 0.393 & 0.385 & 0.327 & 0.029 & 0.294 \\
QRF-TT           & 0.368 & \underline{0.452} & 0.317 & \underline{0.422} & \underline{0.525} & \underline{0.524} & \underline{0.417} & \underline{0.414} & 0.261 & 0.317 \\
QRF-AP           & \underline{0.379} & \underline{0.448} & \underline{0.333} & 0.360 & \underline{0.387} & 0.433 & 0.372 & \underline{0.423} & \underline{0.347} & \underline{0.408} \\
\hline
\rowcolor{gray!20} % Adds a light gray background to the QRF (avg) row
QRF (avg)          & 0.344 & 0.397 & 0.311 & 0.380 & 0.405 & 0.417 & 0.384 & 0.344 & 0.234 & 0.361 \\
\hline
MLP-D-Morgan     & 0.182 & 0.212 & 0.252 & 0.241 & 0.091 & 0.267 & 0.229 & 0.278 & 0.180 & 0.076 \\
MLP-D-MACCS      & 0.237 & 0.281 & 0.273 & 0.217 & 0.204 & 0.299 & 0.225 & 0.076 & 0.210 & 0.184 \\
MLP-D-RDKit      & 0.283 & 0.191 & 0.308 & 0.171 & 0.042 & 0.248 & 0.080 & -0.014 & -0.027 & -0.086 \\
MLP-D-TT         & \underline{0.422} & 0.206 & 0.274 & 0.231 & 0.171 & 0.241 & 0.178 & 0.270 & 0.203 & 0.097 \\
MLP-D-AP         & 0.157 & 0.209 & 0.223 & 0.231 & 0.064 & 0.199 & 0.083 & 0.195 & 0.141 & 0.124 \\
\hline
\rowcolor{gray!20} % Adds a light gray background to the QRF (avg) row
MLP-D (avg)        & 0.256 & 0.220 & 0.266 & 0.218 & 0.114 & 0.251 & 0.159 & 0.188 & 0.141 & 0.079 \\
\hline
% GNN              & {---} & {---} & {---} & {---} & {---} & {---} & {---} & {---} & {---} & {---} \\
GREA             & 0.231 & 0.253 & 0.091 & 0.016 & 0.216 & 0.224 & 0.221 & 0.069 & 0.191 & 0.298 \\
\bottomrule
\end{tabular}
\end{table}

Spearman's rank correlation coefficient ($\rho_s$) measures the strength and direction of the monotonic relationship between prediction errors and uncertainties, which is defined as:

\begin{equation}
\rho_s = 1 - \frac{6 \sum d_i^2}{n(n^2 - 1)},
\label{eq:spearman}
\end{equation}

where $d_i$ represents the difference between the ranks of the prediction errors and their corresponding uncertainties, and $n$ is the total number of data points. A coefficient of 1 indicates a perfect positive monotonic relationship, meaning that higher uncertainties consistently correspond to larger errors. Conversely, a coefficient of $-$1 indicates a perfect negative monotonic relationship, and a coefficient close to 0 suggests no correlation between uncertainty and error. Sparsification plots offer a complementary approach for assessing UQ by systematically removing data points with the highest predicted uncertainties and tracking the cumulative prediction error. The idea is that if the predicted uncertainties are reliable, removing highly uncertain predictions should result in a rapid decline in cumulative error. Therefore, a good UQ model is expected to show a significant reduction in error as more uncertain points are removed. This approach not only provides insight into the quality of the uncertainty estimates but also illustrates how model performance, in terms of both accuracy and reliability, improves when uncertain predictions are excluded. Details on the calculation of prediction uncertainties are available in Section \ref{sec:MLmodel}.

\begin{figure}[tpb]
    \centering
    \includegraphics[width=1\linewidth]{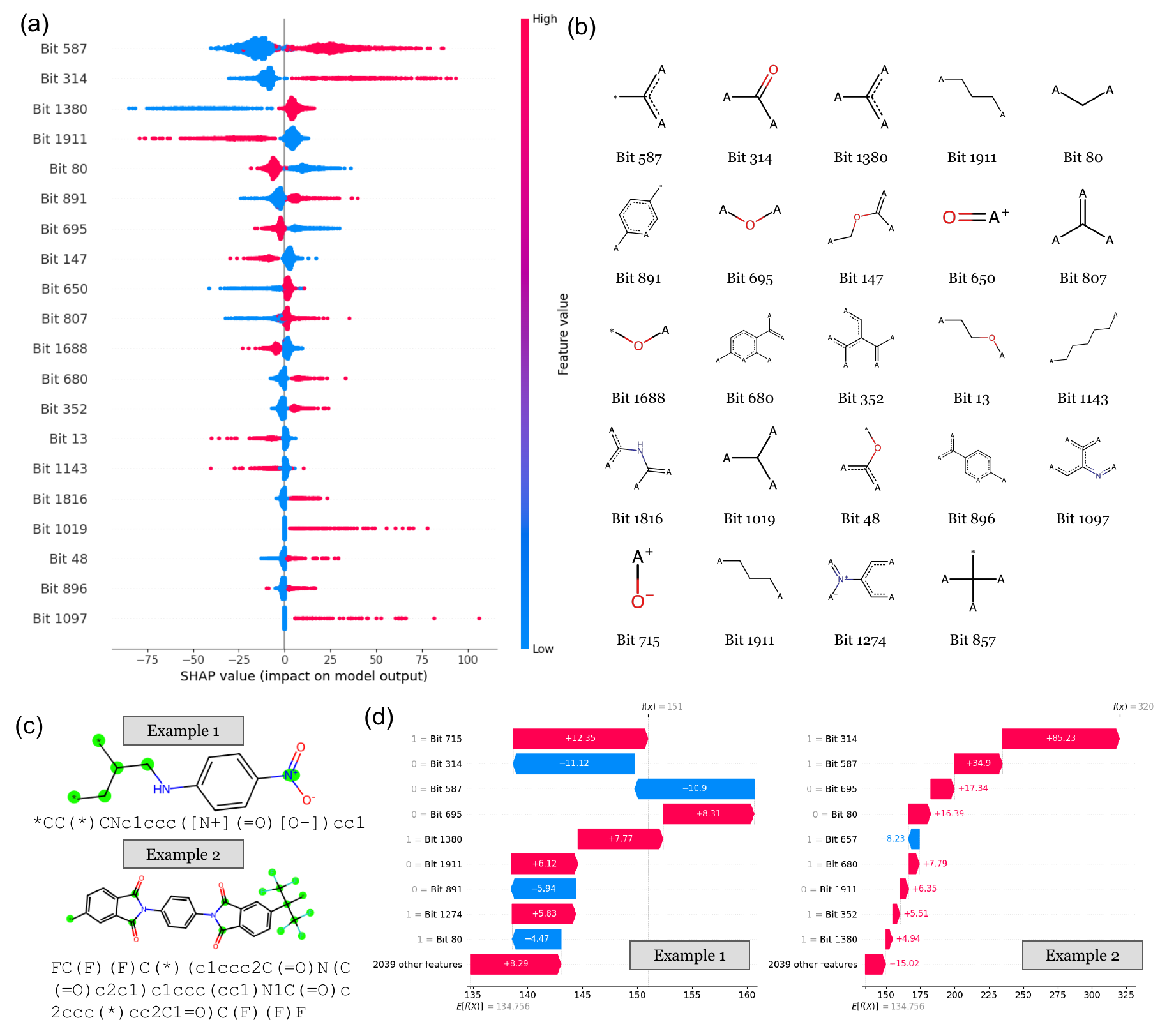}
    % \vspace{-0.1in}
    \caption{An example of model prediction interpretation on the T\textsubscript{g} test dataset. (a) Beeswarm plot of SHAP values on the test dataset using the QFR model and Morgan fingerprints. The x-axis represents the SHAP values, which quantify the impact of each fingerprint bit on the model's prediction—positive values increase T\textsubscript{g}, while negative values decrease it. The y-axis lists the top-20 most important fingerprint bits, ranked in descending order by their average absolute SHAP value (i.e., the most influential bits are at the top). The color of the dots corresponds to the feature value: red indicates bit=1 in Morgan fingerprints, while blue represents bit=0. (b) Molecular visualization of important bits in the Morgan fingerprint. ``A'' is a wildcard atom represents any atom type and ``*'' represents the polymerization point in the repeated unit of polymers. (c) Molecular structure and rationale interpretation (highlighted in green) of two polymer explicands from the GREA model. (d) Waterfall plot of SHAP values of the same two polymer explicands in panel (c) using the QFR model and Morgan fingerprints. The x-axis shows the cumulative SHAP value contributions leading to the final prediction $f(x)$, with the base value (expected model output) at the far left and the final model prediction at the far right. Each bar represents the contribution of a single fingerprint bit. Bars are annotated with the bit ID and the magnitude of their contribution.}
    \label{fig:explain_Tg}
    % \vspace{-0.25in}
\end{figure}

The results of $\rho_s$, summarized in Table \ref{tab:test_uncertain}, highlight distinct differences in UQ performance across various models. Despite its lower predictive accuracy observed in Table \ref{tab:test_acc_comparison}, QRF showed the best UQ capability, with $\rho_s$ values generally between 0.3 and 0.5—typically considered moderate for molecular and polymer property prediction~\cite{hirschfeld2020uncertainty, tang2024uncertainty}. This can be attributed to its intrinsic mechanism of quantile estimation, which models the conditional distribution of the target variable, enabling it to provide more reliable uncertainty estimates. Although QRF may not excel in predicting the exact value of a property, it can more effectively quantify the confidence level of its predictions, making it suitable for scenarios where uncertainty estimation is important. MLP-D and GREA exhibited lower UQ performance. Although MLP-D uses dropout regularization to approximate Bayesian inference by simulating model uncertainty through stochastic forward passes, its UQ performance was less robust than QRF. This is likely because, while dropout helps reduce overfitting and improve generalization, it does not explicitly model the full distribution of the target variable. GREA’s UQ strength lies in its rationale-environment separation mechanism, which identifies key subgraphs (rationales) responsible for predictions. This enhances interpretability and provides a degree of uncertainty estimation by assessing the variability of rationales across different environments. However, unlike QRF, the environment replacement strategy in GREA does not offer an explicit estimate of the full distribution of the target variable, which also limits its UQ precision. The sparsification plots of the models with the best $\rho_s$ for each property are shown in Fig. \ref{fig:test_parity}, which illustrate how prediction error and prediction uncertainty accumulate as samples with the highest uncertainties are sequentially removed. In these plots, the models show a rapid decrease in cumulative prediction error as the most uncertain samples are excluded, demonstrating that the predicted uncertainties effectively capture regions of high model error. 

% The choice of polymer fingerprints also appeared to influence UQ performance. Morgan fingerprints, which capture detailed local structural information of polymers, tended to facilitate better UQ compared to simpler fingerprints such as Topological fingerprints. This suggests that fingerprints capable of representing subtle variations in polymer structure enable models to better estimate uncertainty. Additionally, the complexity of the target property affects UQ effectiveness. For properties like density ($\rho$) and glass transition temperature (T\textsubscript{g}), which are closely tied to the polymer’s intrinsic chemical structure, simpler models and representations may suffice for reliable UQ. However, for more complex properties such as thermal conductivity (TC) and gas permeability (P), which depend on both local and global interactions, more advanced models and representations are needed to achieve accurate uncertainty quantification.

% These findings emphasize that while prediction accuracy remains a key objective, reliable uncertainty quantification is equally important for robust predictive modeling in polymer informatics. By leveraging models and representations that balance accuracy and UQ capability, researchers can improve both the reliability and practical applicability of predictive tools in polymer discovery.

\subsection{Interpretability}

Understanding model predictions from both global and local perspectives is essential for improving model reliability, uncovering underlying patterns in the data, and facilitating decision-making in downstream tasks. Global interpretability helps identify key features that consistently influence predictions, enabling researchers to discover potential underlying mechanisms or general patterns across the dataset. Local interpretability, on the other hand, provides detailed insights into how specific features of individual data points contribute to a prediction, which is crucial for selecting candidates for experimental validation and further investigation. Both levels of interpretability are important for increasing trust in ML models and ensuring that predictions are not only accurate but also explainable.

For global interpretability, we used SHAP (details provided in Section \ref{sec:method_interpretability}) to calculate feature importance and identify key bits in the fingerprints that influence predictions. Figure \ref{fig:explain_Tg}(a) shows a beeswarm plot of SHAP values for the QRF model on the T\textsubscript{g} test dataset using Morgan fingerprints. Each bit corresponds to a specific molecular structure, as listed in Fig. \ref{fig:explain_Tg}(b). Bits 587, 314, and 1380 have the most positive impact on T\textsubscript{g}, while bits 1911 and 80 have the most negative impact. A rigid aromatic structure on the backbone (with the aromatic ring in Bit 587 directly connected to a polymerization point "*") or an aromatic structure at an unspecified position (Bit 1380) positively influences T\textsubscript{g}, consistent with the general understanding that rigid, planar structures increase T\textsubscript{g} by restricting chain mobility~\cite{liu2010prediction}. Similarly, Bit 314, representing a carbonyl group, positively influences T\textsubscript{g} by enhancing intermolecular interactions and reducing chain mobility~\cite{yu2017quantitative}. On the other hand, Bit 1911 and Bit 80, which represent flexible aliphatic chains, negatively influence T\textsubscript{g} by increasing chain mobility. Bit 1911 corresponds to a longer linear alkyl chain, while Bit 80 represents a shorter alkyl linkage. These flexible, non-polar structures reduce intermolecular interactions and allow easier interchain movements, which lower T\textsubscript{g}~\cite{liu2010prediction,yu2017quantitative}.

For local interpretability, we utilized two approaches: one for graph-based models and another for non-graph-based models. The GREA model, being a graph-based approach, provides inherent rationale interpretation by identifying atom-level importance. Figure \ref{fig:explain_Tg}(c) illustrates two examples from the T\textsubscript{g} dataset, where the important nodes are highlighted in green, indicating the substructures that play a critical role in the property prediction, i.e., the rationales. For non-graph-based models, we again used SHAP for local explanations. The explanation of the two same polymers from the T\textsubscript{g} dataset is shown in Fig. \ref{fig:explain_Tg}(d), where the bit-level contributions to the prediction are visualized using a waterfall plot, with their corresponding molecular structures listed in Fig. \ref{fig:explain_Tg}(b). Comparing the local explanations from graph-based and non-graph-based models reveals that both approaches identify similar key structural elements. For instance, in example 1, both methods highlight the significance of the nitro group and alkyl linkages, while in example 2, they both emphasize the importance of carbonyl groups and tertiary carbon centers. The graph-based GREA model offers atom-level interpretability, providing chemists with a direct understanding of specific functional groups. In contrast, the SHAP-based approach for non-graph models delivers bit-level importance, which, although less granular, is useful for identifying broader structural patterns and offers insights into whether certain features have a positive or negative impact, which is not provided by GREA. By combining insights from both approaches, we can gain a more comprehensive understanding of the factors influencing polymer property predictions, ultimately improving model transparency and trustworthiness.

\subsection{Synthesizability}
We manually curated a dataset of polymerization reactions from literature and various resources \cite{chen2021data, SPEIGHT2011499, SHRIVASTAVA201817, carothers1931polymerization, odian2004principles, aoi1996polymerization, mandal2012fundamentals, liaw2012advanced, sroog1991polyimides, lin2005recent}, with a focus exclusively on linear homopolymers with two polymerization points, omitting complex structures like ladder polymers and copolymers. In this study, we considered three primary polymerization types: condensation, addition, and ring-opening polymerization. The curated reaction data contains 578 polymerization reactions, which were randomly split into 218 training and 360 testing samples. Each reaction comprises (1) the polymer’s SMILES (Simplified Molecular Input Line Entry System) \cite{weininger1988smiles}, (2) the polymerization type, and (3) the SMILES of the corresponding monomer(s).  All the monomers involved in these reactions can be found in PubChem~\cite{kim2019pubchem}, which suggests that they are commercially available or can be synthesized. Using the training set, we defined a set of reaction templates based on SMARTS (SMILES Arbitrary Target Specification)~\cite{daylight_smarts} to retro-synthetically decompose polymers into their respective monomers. 

\begin{figure}[tpb]
    \centering
    \includegraphics[width=0.7\linewidth]{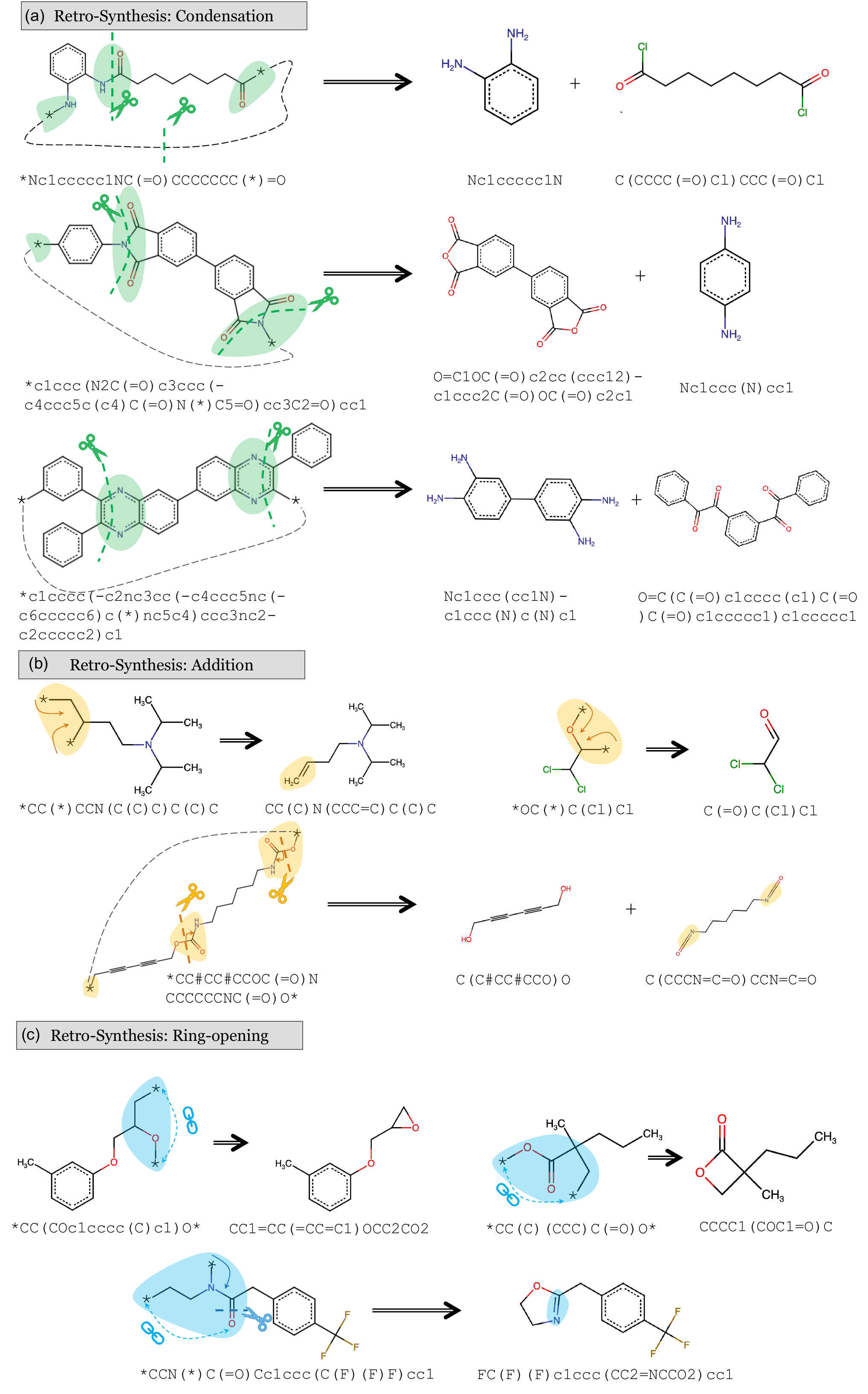}
    % \vspace{-0.1in}
    \caption{Examples of retrosynthesis planning of polymers from: (a) condensation, (b) addition, and (c) ring-opening polymerization. The condensation reactions involve monomers with reactive groups that release small molecules upon polymerization. Addition polymerizations involve linking monomers via reactive double bonds. Ring-opening polymerizations involve the cleavage of ring structures to form linear chains.}
    \label{fig:retro-all}
    % \vspace{-0.25in}
\end{figure}

Examples of the retro-synthesis procedure using these templates are illustrated in Fig. \ref{fig:retro-all}. Condensation polymerization follows a step-growth mechanism, where a small molecule byproduct (such as water or HCl) is released during the polymerization process. Figure \ref{fig:retro-all}(a) illustrates the retrosynthetic pathways for typical condensation reactions, demonstrating how polymers can be deconstructed into their monomeric units (byproducts are omitted). Addition polymerization proceeds through a chain-growth mechanism without the release of byproducts. Figure \ref{fig:retro-all}(b) depicts the retrosynthetic pathways for addition polymerizations, illustrating the deconstruction of polymers into their respective unsaturated monomer units. Ring-opening polymerization involves the opening of a ring structure in the monomer to form a linear polymer chain. Examples of the corresponding retrosynthetic pathways are shown in Fig. \ref{fig:retro-all}(c).

\begin{table}[ht]
\centering
\caption{Summary of Training and Testing Datasets and Evaluation Metrics for Polymerization Reaction Templates.}
\label{tab:polymerization_metrics}
\begin{tabular}{lcc}
\toprule
\textbf{Dataset}     & \textbf{Prediction Accuracy (\%)} & \textbf{Applicability Rate (\%)} \\
\midrule
Training             & 100                        & 100                                \\
Testing              & 59.7                        & 91.9                                \\
\bottomrule
\end{tabular}
\end{table}

We extracted 82 templates from the training set and evaluated their performance on the hold-out test set. Two metrics were employed for this evaluation: 
\begin{enumerate}
    \item \textit{Prediction Accuracy}: This metric assesses whether the ground truth polymerization reaction is accurately predicted by any of the existing templates. For each polymer in the dataset, a score of 1 is assigned if the ground truth reaction is among the predicted results; otherwise, a score of 0 is given. The Prediction Accuracy is then the percentage of polymers with a score of 1 across the dataset.
    \item \textit{Applicability Rate}: This metric evaluates the applicability of existing templates to a given polymer. For each polymer, a score of 1 is assigned if at least one prediction result is generated using the current templates; if no predictions are possible due to limitations in template coverage, a score of 0 is assigned. The Applicability Rate is the percentage of polymers with a score of 1 in the dataset.
\end{enumerate}

Table \ref{tab:polymerization_metrics} presents the results of the evaluation metrics for the polymerization reaction templates applied to both the training and testing datasets. The Prediction Accuracy and Applicability Rate for the training set are both 100\%, which reflects the fact that the templates were specifically designed and tailored to cover all the reactions in the training data. Therefore, every reaction in the training set is correctly predicted by the templates. For the testing dataset, the Applicability Rate remains high at 91.9\%, indicating that the existing templates cover a large portion of the potential polymer linkages present in the test set. This demonstrates the broad applicability of the templates to various polymerization reactions, even those not explicitly included in the training data. However, the Prediction Accuracy for the testing set is 59.7\%. This does not imply that the predictions for the testing data are incorrect, but rather that the reaction routes for some polymers in the test set cannot be fully covered by the current templates. It is important to note that one polymer may have multiple "correct" reaction routes, and the templates may not capture all possible pathways. As more templates are extracted, the accuracy is expected to improve by covering a broader range of reaction possibilities.

Based on the retrosynthesis process, we propose a new synthesizability score specific to polymers, named \texttt{PolyScore}. Unlike existing retrosynthesis scores designed for small molecules, PolyScore is tailored to capture the unique characteristics of polymerization reactions, which considers the difficulty of synthesizing a polymer by evaluating the complexity of its retro-synthesized monomers and the number of viable polymerization routes. The PolyScore (\(S_{\text{Poly}}\)) is defined as:

\begin{equation}
S_{\text{Poly}} = \min\left(S_{\text{Poly}}^j\right), \quad S_{\text{Poly}}^j = \text{Mean}_{\text{Reactant}} \left( S_{\text{Reactant}}^{j} \right) = \frac{n}{\sum_{i=1}^{n} \frac{1}{S_{\text{Reactant}, i}^j}} 
\label{eq:polyscore}
\end{equation}

where $S_{\text{Reactant}, i}^j$ represents the synthetic accessibility score (SAScore~\cite{ertl2009estimation}) of the $i^{th}$ retro-synthesized monomer in the $j^{th}$ route. The final PolyScore ($S_{\text{Poly}}$) is taken as the minimum of the scores across all possible polymerization routes ($S_{\text{Poly}}^j$). This approach ensures that polymers with easier or more feasible synthesis routes receive a higher score, reflecting their higher likelihood of successful synthesis in practice. However, the current version of PolyScore has certain limitations. It only considers the synthesizability of monomers when a polymer can be successfully decomposed into chemically valid monomers based on our polymerization retrosynthesis templates. Factors such as the reaction conditions (e.g., temperature, pressure, solvents, and catalysts) that can impact the feasibility of polymerization are not accounted for. Additionally, the influence of side reactions, commercial availability of monomers, and industrial-scale synthetic constraints is not accounted for. Future enhancements to PolyScore could incorporate these aspects to provide a more comprehensive measure of polymer synthesizability.

% In summary, by combining a systematic dataset of polymerization reaction templates with a novel synthesizability score, we provide a framework for evaluating the feasibility of synthesizing polymers. PolyScore, in particular, offers a valuable tool for researchers working on generative polymer models, enabling them to assess and prioritize candidate polymers based on their synthetic accessibility.

% \begin{figure}
%     \centering
%     \includegraphics[width=0.72\linewidth]{condensation.pdf}
%     % \vspace{-0.1in}
%     \caption{Examples of retrosynthesis planning of polymers from condensation polymerization.}
%     \label{fig:condensation}
%     % \vspace{-0.25in}
% \end{figure}

% \begin{figure}
%     \centering
%     \includegraphics[width=0.75\linewidth]{addition.pdf}
%     % \vspace{-0.1in}
%     \caption{Examples of retrosynthesis planning of polymers from addition polymerization.}
%     \label{fig:addition}
%     % \vspace{-0.25in}
% \end{figure}

% \begin{figure}
%     \centering
%     \includegraphics[width=0.6\linewidth]{ring.pdf}
%     % \vspace{-0.1in}
%     \caption{Examples of retrosynthesis planning of polymers from ring-opening polymerization.}
%     \label{fig:ring}
%     % \vspace{-0.25in}
% \end{figure}

\subsection{Screening}

In this section, we demonstrate the use of trained models to screen an unlabeled or virtual polymer database, identify promising candidates for specific applications, and apply the retro-synthesis tool to propose polymerization routes. The PI1M \cite{ma2020pi1m} virtual polymer database was selected as the screening pool. We conducted two case studies to illustrate how the screening and retro-synthesis workflow can be potentially utilized for real-world applications: (1) designing high-performance polymers for thermal management materials and (2) developing polymers for gas separation membranes. A complete set of screening results on the PI1M database, including predicted properties, uncertainties, and suggested polymerization routes, can be found at \url{https://github.com/Jiaxin-Xu/POINT2.git}. It should be noted that in the case studies, some properties, such as TC and FFV, are derived from computational simulations rather than experimental sources. As a result, the magnitudes used here might differ from experimental intuitions due to systematic differences between computational simulations and experimental measurements \cite{tao2023machine, ma2022machine}.

\subsubsection*{Case Study 1: Designing a High-Performance Polymer for Thermal Management}

\begin{table}[h!]
\centering
\caption{Property Constraints and Rationale for Case Study 1 - Designing a High-Performance Polymer for Thermal Management. Target ranges are defined for T\textsubscript{g}, T\textsubscript{m}, thermal conductivity (TC), density ($\rho$), and fractional free volume (FFV) to ensure thermal stability, heat dissipation, and mechanical reliability. The ranges for TC and FFV are based on MD simulation labels, which may slightly overestimate values relative to experimental data\cite{tao2023machine, ma2022machine}.}
\label{tab:case_study_1}
\renewcommand{\arraystretch}{1.5}  % Adjust row height
\begin{tabular}{|C{2cm}|C{3cm}|p{8cm}|}
\hline
\textbf{Property} & \textbf{Range} & \textbf{Rationale} \\
\hline
T\textsubscript{g} & $>$ 250°C & Ensures thermal stability; the polymer remains in its glassy state and maintains its mechanical integrity under high operational temperatures encountered in thermal management applications. \\
\hline
T\textsubscript{m} & $>$ 350°C & Prevents melting or deformation under high temperature conditions, ensuring long-term reliability in heat-intensive environments. \\
\hline
TC & $>$ 0.35 W/mK & High thermal conductivity facilitates effective heat dissipation, which is essential for preventing overheating in electronic devices and other applications. \\
\hline
$\rho$ & 0.8 – 1.2 g/cm$^3$ & Ensures a balance between mechanical robustness and lightweight properties, aiding in ease of integration into various devices while minimizing added weight. \\
\hline
FFV & 0.3 - 0.35 & Low fractional free volume promotes tight chain packing in the polymer matrix, which reduces phonon scattering and energy barriers for thermal conduction. \\
\hline
\end{tabular}
\end{table}

\begin{figure}[tpb]
    \centering
    \includegraphics[width=1\linewidth]{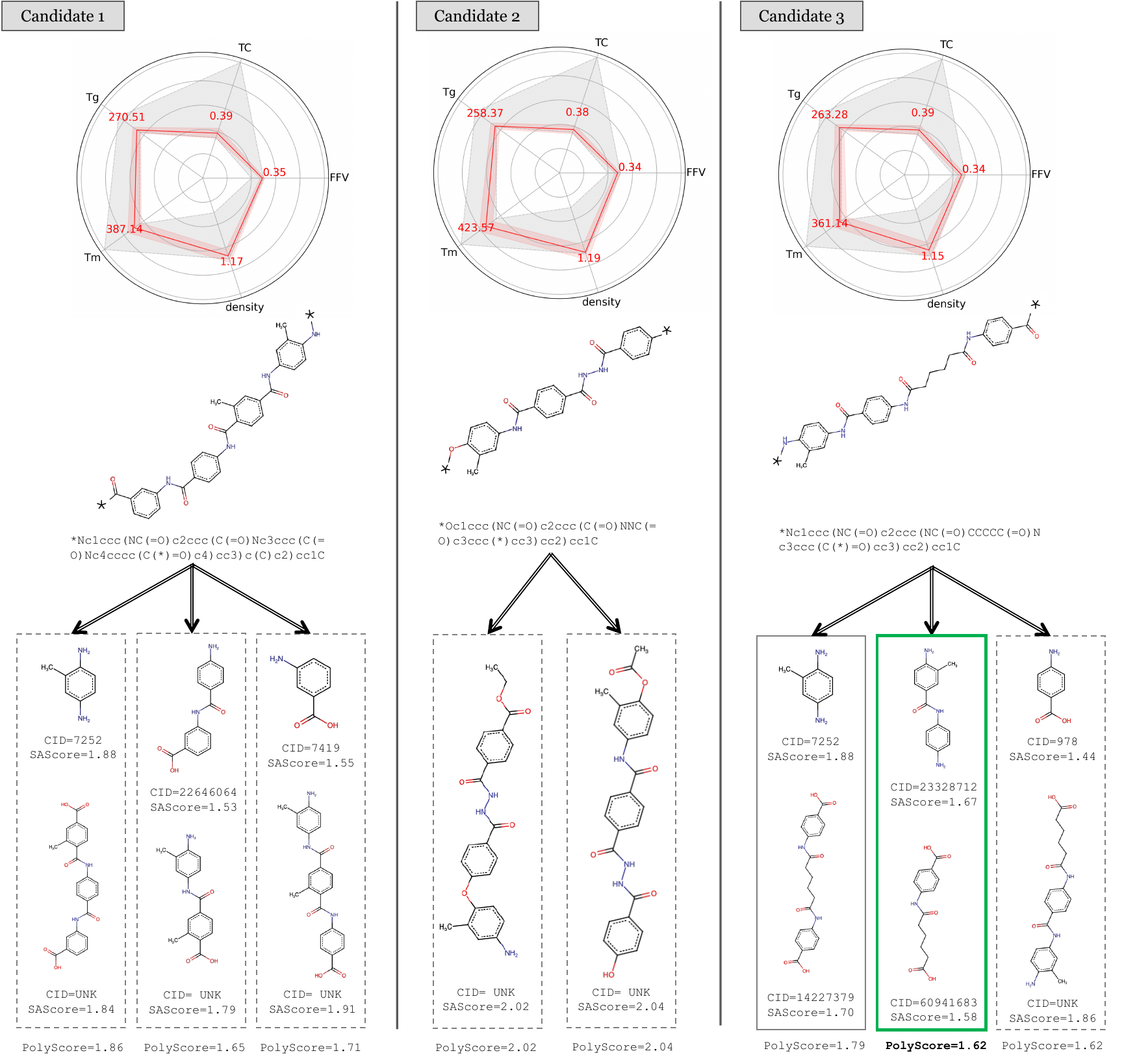}
    % \vspace{-0.1in}
    \caption{Results of Case Study 1: designing high-performance polymers for thermal management. The top row shows radar plots with design constraints (shaded gray regions) for key properties: density (\(0.8{-}1.2 \, \mathrm{g/cm^3}\)), FFV (\(0.3{-}0.35\)), TC (\(>0.35\, \mathrm{W/mK}\)),  T\textsubscript{g} (\(> 250^\circ\mathrm{C}\)), and  T\textsubscript{m} (\(>350^\circ\mathrm{C}\)). The red line and numbers indicate the predicted mean property values, while the shaded red area represents the uncertainty range. Models used for predictions are: FFV (MLP-D+AP), TC (MLP-D+TT), T\textsubscript{g} (MLP-D+Morgan),T\textsubscript{m} (MLP-D+RDKit), and density (MLP-D+Morgan). The middle row displays the molecular structure and SMILES of the candidate polymers. The bottom row shows retrosynthetic pathways, where each box represents a potential route. Boxes with solid borders indicate all monomers are available in PubChem (CID provided), while dashed borders denote at least one monomer is unknown (UNK). The synthesizability of monomers is quantified by SAScore. The PolyScore of each proposed route is shown at the bottom. Candidate 3 is the final selection due to its optimal properties and the most feasible synthesis route, as highlighted with a green box.
}
    \label{fig:case-1}
    % \vspace{-0.25in}
\end{figure}

In the first case study, the objective is to design a high-performance polymer for thermal management applications, such as heat dissipation in electronic devices, batteries and aerospace components. The polymer must meet specific property requirements of high thermal conductivity, mechanical integrity, and high thermal stability during operation \cite{zhang2021recent,vadivelu2016polymer}. Table \ref{tab:case_study_1} summarizes these property requirements along with their design rationale. Utilizing selected models that consider both prediction accuracy and UQ, we screened a randomly sampled 10\% subset of the PI1M database and identified three polymer candidates that satisfied all the specified property requirements. The predicted properties and retrosynthetic routes of the three candidates are shown in Fig. \ref{fig:case-1}. Among the candidates that meet the design requirements, Candidate 3, classified as a polyamide, was selected as the final choice due to its synthetic feasibility. This polymer can be synthesized via polycondensation using readily available monomers from PubChem \cite{kim2019pubchem}, including CID=23328712 (4-amino-N-(4-aminophenyl)-3-methylbenzamide) and CID=60941683 (4-(5-carboxypentanoylamino)benzoic acid). The retrosynthetic pathway, emphasizing these monomers, is depicted in the green box of Fig. \ref{fig:case-1}. Although another pathway consisting solely of known monomers from PubChem is feasible, the higher SAScores \cite{ertl2009estimation} of the monomers indicate greater synthesis difficulty. Thus, the route with the easiest synthesis process was selected as the final choice, which was further confirmed by the lowest PolyScore of the route.

\subsubsection*{Case Study 2: Designing a High-Performance Polymer for Gas Separation Membranes}

\begin{table}[h!]
\centering
\caption{Property Constraints and Rationale for Case Study 2 - Designing a High-Performance Polymer for CO$_2$/CH$_4$ Gas Separation Membranes. Target ranges are defined for T\textsubscript{g}, density ($\rho$), fractional free volume (FFV), and log$_{10}$-scaled CO$_2$ and CH$_4$ gas permeability in units of Barrer to ensure structural integrity, thermal stability, and selective gas transport. The ranges for FFV are based on MD simulation labels, which may slightly overestimate values relative to experimental data\cite{tao2023machine}.}
\label{tab:case_study_2}
\renewcommand{\arraystretch}{1.5}  % Adjust row height
\begin{tabular}{|C{3cm}|C{2cm}|p{8cm}|}
\hline
\textbf{Property} & \textbf{Range} & \textbf{Rationale} \\
\hline
T\textsubscript{g} & $>$ 180°C & Ensures good thermal stability and rigid polymer backbone for size sieving. \\
\hline
% T\textsubscript{m} & $>$ 300°C & Prevents melting under high operational temperatures, ensuring the polymer retains its structural form and functionality. \\
% \hline
% TC & $>$ 0.5 W/mK & Facilitates efficient heat dissipation, preventing localized overheating and potential degradation of membrane performance. \\
% \hline
FFV & $>$ 0.35 & Provides sufficient free space within the polymer matrix to facilitate gas diffusion. \\
\hline
$\rho$ & $>$ 1.4 g/cm$^3$ & High density ensures structural integrity under operational pressures, reducing the risk of polymer deformation or failure in demanding gas separation environments. \\
\hline
log$_{10}$(P\textsubscript{CO$_2$} in Barrer) & $>$ 1.5 & High permeability to CO$_2$ enhances separation efficiency, crucial for applications like carbon capture and natural gas purification. \\
\hline
log$_{10}$(P\textsubscript{CH$_4$} in Barrer) & $<$ -0.8 & Low permeability to CH$_4$ ensures methane is effectively retained, improving product purity and process efficiency. \\
\hline
\end{tabular}
\end{table}

\begin{figure}[tpb]
    \centering
    \includegraphics[width=1\linewidth]{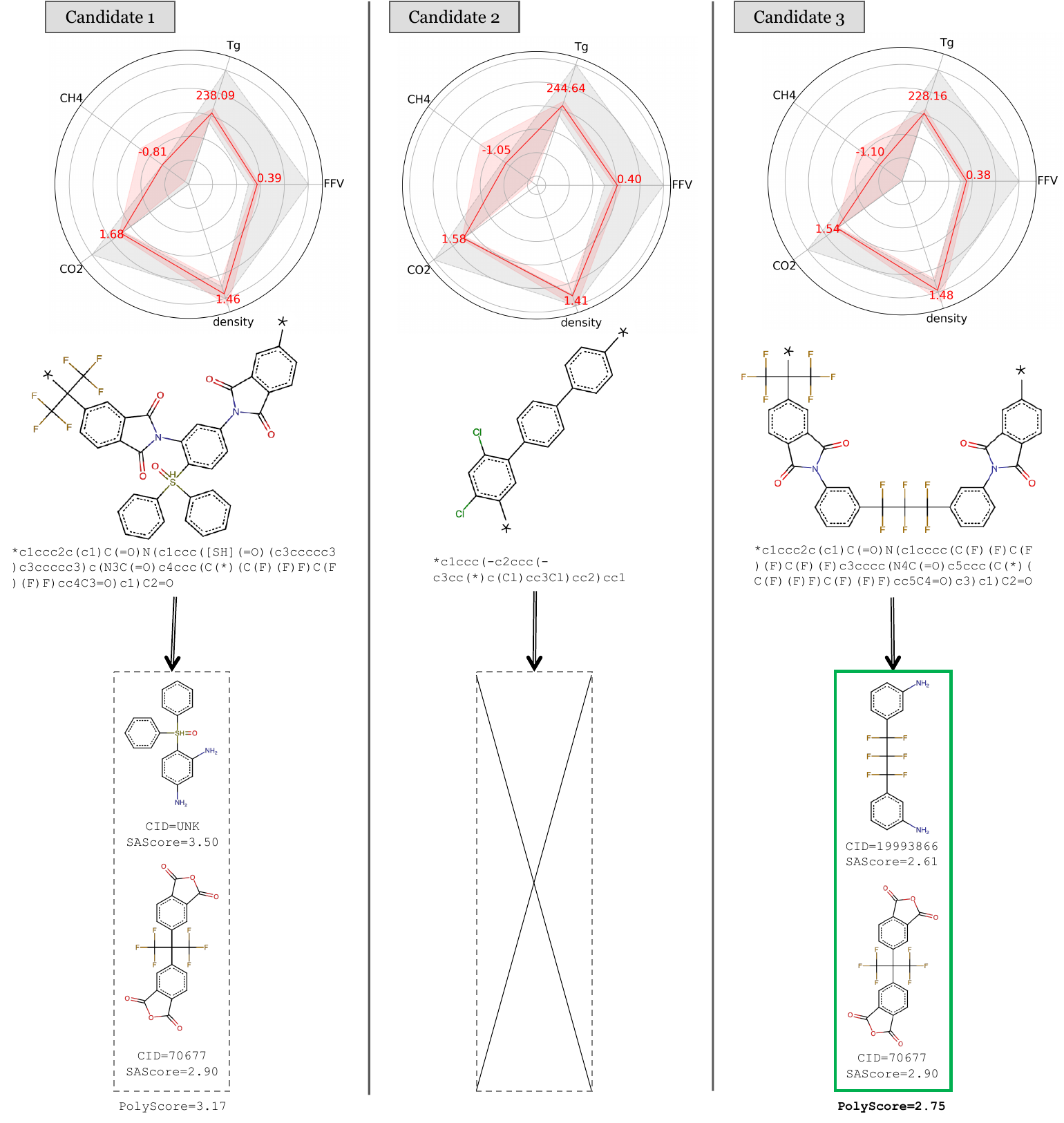}
    % \vspace{-0.1in}
    \caption{Results of Case Study 2: designing high-performance polymers for gas separation membranes. The top row shows radar plots with design constraints (shaded gray regions) for key properties: density (\(>1.4 \, \mathrm{g/cm^3}\)), FFV (\(>0.35\)), T\textsubscript{g} (\(>180^\circ\mathrm{C}\)), log$_{10}$(P\textsubscript{CH$_4$} in Barrer) (\(<-0.8\)), and log$_{10}$(P\textsubscript{CO$_2$} in Barrer) (\(>1.5\)). The red line and numbers indicate the predicted mean property values, while the shaded red area represents the uncertainty range. Models used for predictions are: FFV (MLP-D+AP), T\textsubscript{g} (MLP-D+Morgan), density (MLP-D+Morgan), P(CH$_4$) (QRF+TT), and P(CO$_2$) ((MLP-D+AP)). The middle row displays the molecular structure and SMILES of the candidate polymers. The bottom row shows retrosynthetic pathways, where each box represents a potential route. No route is identified using current templates for Candidate 2. Boxes with solid borders indicate all monomers are available in PubChem (CID provided), while dashed borders denote at least one monomer is unknown (UNK). The synthesizability of monomers is quantified by SAScore. The PolyScore of each proposed route is shown at the bottom. Candidate 3 is the final selection due to its optimal properties and the most feasible synthesis route, as highlighted with a green box.}
    \label{fig:case-2}
    % \vspace{-0.25in}
\end{figure}

In the second case, the objective is to design a high-performance polymer for CO$_2$/CH$_4$ gas separation membranes, particularly for applications such as natural gas purification. The polymer must meet specific property requirements to ensure efficiency, durability, and selectivity during operation \cite{sanders2013energy}. Table \ref{tab:case_study_2} summarizes these property constraints along with their application rationale. Using the trained ML models and screening workflow, three candidates were identified as shown in Fig. \ref{fig:case-2}. Among them, Candidate 3 was selected as the final choice based on its synthetic feasibility and suitability for the target CO$_2$/CH$_4$ separation application. Candidate 3, a polyimide, is a particularly promising choice as polyimides are widely recognized in gas separation membrane design for their excellent thermal stability, mechanical strength, and tunable chemical structures, which allow for tailored permeability and selectivity \cite{sanders2013energy}. The synthesis route for Candidate 3 is highlighted in the green box in Fig. \ref{fig:case-2}. This polymer can be synthesized via polycondesation using two known monomers from PubChem: 3,3'-(Hexafluorotrimethylene)bisaniline (CID=19993866) and 4,4'-(Hexafluoroisopropylidene)diphthalic anhydride (CID=70677).

% \cite{kour2014real,kour2014fast} and see \cite{hadash2018estimate}.

% The documentation for \verb+natbib+ may be found at
% \begin{center}
%   \url{http://mirrors.ctan.org/macros/latex/contrib/natbib/natnotes.pdf}
% \end{center}
% Of note is the command \verb+\citet+, which produces citations
% appropriate for use in inline text.  For example,
% \begin{verbatim}
%    \citet{hasselmo} investigated\dots
% \end{verbatim}
% produces
% \begin{quote}
%   Hasselmo, et al.\ (1995) investigated\dots
% \end{quote}

% \begin{center}
%   \url{https://www.ctan.org/pkg/booktabs}
% \end{center}

% \subsection{Figures}

% See Figure \ref{fig:fig1}. Here is how you add footnotes. \footnote{Sample of the first footnote.}
% \lipsum[11] 

% \begin{figure}
%   \centering
%   \fbox{\rule[-.5cm]{4cm}{4cm} \rule[-.5cm]{4cm}{0cm}}
%   \caption{Sample figure caption.}
%   \label{fig:fig1}
% \end{figure}

% \subsection{Tables}

% See awesome Table~\ref{tab:table}.

% \begin{table}
%  \caption{Sample table title}
%   \centering
%   \begin{tabular}{lll}
%     \toprule
%     \multicolumn{2}{c}{Part}                   \\
%     \cmidrule(r){1-2}
%     Name     & Description     & Size ($\mu$m) \\
%     \midrule
%     Dendrite & Input terminal  & $\sim$100     \\
%     Axon     & Output terminal & $\sim$10      \\
%     Soma     & Cell body       & up to $10^6$  \\
%     \bottomrule
%   \end{tabular}
%   \label{tab:table}
% \end{table}

% \subsection{Lists}
% \begin{itemize}
% \item Lorem ipsum dolor sit amet
% \item consectetur adipiscing elit. 
% \item Aliquam dignissim blandit est, in dictum tortor gravida eget. In ac rutrum magna.
% \end{itemize}

\section{Conclusion}
In this study, we introduced POINT$^2$, a comprehensive polymer informatics framework that integrates property prediction, uncertainty quantification, interpretability, and synthesizability to facilitate the design and discovery of high-performance polymers. By leveraging advanced ML models and diverse polymer representations, we demonstrated the ability to predict key polymer properties, quantify uncertainties, interpret prediction results, and propose feasible polymerization routes using a template-based retrosynthesis tool. The introduction of PolyScore provides a polymer-specific synthesizability metric, enabling practical assessments of the ease of polymer synthesis. Through two case studies, we showcased the potential application of POINT$^2$, identifying and selecting polymers tailored for thermal management and gas separation membranes. These case studies highlight the utility of combining property screening with retrosynthetic analysis to balance predictive performance and synthetic feasibility.

We envision POINT$^2$ as a continually evolving framework, with future developments aimed at enhancing its capabilities and broadening its applicability. In the near term, efforts will focus on addressing current limitations. For example, the current PolyScore focuses exclusively on monomer complexity and route viability without considering reaction conditions, side reactions, or industrial-scale constraints. Additionally, while UQ provides valuable insights into prediction reliability, improving its calibration across diverse property spaces is still challenging. This involves addressing both aleatoric uncertainty, which arises from noise and variability in data (e.g., gas permeability or thermal conductivity), and epistemic uncertainty, which stems from the model’s limited knowledge, especially in sparsely sampled regions of the polymer chemical space. Future enhancements could incorporate experimental validation of predicted results, expanded polymerization templates, and hybrid data-driven approaches to better capture polymer-specific complexities.

We believe POINT$^2$ establishes a robust foundation for polymer informatics and will serve as a valuable resource for the broader community, advancing both theoretical understanding and practical applications of polymer discovery.

\section{Method}
\subsection{Polymer Representations}
\label{sec:representation}
In this study, we employed various molecular fingerprinting techniques to represent polymer structures from SMILES \cite{weininger1988smiles} to numerical values, each capturing distinct aspects of molecular information:

\textbf{Morgan Fingerprints}: Also known as circular fingerprints, these are generated using the Morgan algorithm, which iteratively encodes atomic environments up to a specified radius \cite{rogers2010extended}. This method effectively captures structural information and is widely used for similarity assessments and ML applications \cite{glen2006circular,zhong2023count,pattanaik2020molecular}. The radius parameter determines the neighborhood size considered around each atom, and the bit vector size defines the length of the fingerprint. Commonly used parameters, a radius of 2 and a bit vector size of 2048, were used in this work through RDKit \cite{landrum2013rdkit}.

\textbf{MACCS Keys}: The MACCS (Molecular ACCess System) keys \cite{durant2002reoptimization} comprise a set of 166 predefined structural fragments used to represent molecular structures as binary vectors, where each bit indicates the presence or absence of a specific substructure. In this work, we utilized RDKit \cite{landrum2013rdkit} to generate MACCS fingerprints for our dataset. 

\textbf{RDKit Fingerprints}: RDKit \cite{landrum2013rdkit} provides topological fingerprints that encode the presence of various substructures within a molecule, facilitating substructure searches and similarity comparisons. In this work, we configured the RDKit fingerprints with a bit vector size of 2048 bits, a minimum path length of 1 bond, a maximum path length of 6 bonds, and 2 bits set per hash function. 

\textbf{Atom Pair Fingerprints}: These fingerprints capture information about all pairs of atoms in a molecule, encoding their atom types and the topological distance between them, valuable for understanding molecular geometry and is often used in similarity assessments \cite{carhart1985atom}. A bit vector size of 2048 and 4 bits per entry were used in this work.

\textbf{Topological Torsion Fingerprints}: Focusing on sequences of four connected atoms, these fingerprints capture information about their types and connectivity, which are particularly useful for characterizing conformational aspects of molecules \cite{nilakantan1987topological}. A bit vector size of 2048 and a target size (the number of atoms in the torsion) of 4 were used in this work.

\textbf{Graph-Based Descriptors}: Utilizing RDKit \cite{landrum2013rdkit}, we extracted graph-based descriptors that represent molecular structures as graphs, with atoms as nodes and bonds as edges. This approach allows for the capture of complex structural information, facilitating the application of graph-based ML models. We defined nine categories for node features, namely (1) atomic number, (2) degree of the atom, (3) chirality, (4) formal charge, (5) total number of $Hs$ (explicit and implicit) on the atom, (6) number of radical electrons, (7) hybridization, (8) aromaticity, and (9) is in a ring or not. Three categories for edge features are defined, namely (1) bond type, (2) is conjugated or not , and (3) stereo configuration. Each polymer is treated as an undirected graph \(G = (X;E;A)\), where \(X\) is the node feature matrix, \(E\) is the edge feature
matrix, and \(A\) is the adjacency matrix.

\subsection{Machine Learning Models}
\label{sec:MLmodel}
We employed several ML models to predict polymer properties considering prediction accuracy, uncertainty estimation, and interpretability:

\textbf{Quantile Random Forests (QRF)}: It extends the traditional random forest algorithm to estimate conditional quantiles, offering a non-parametric method to model the conditional distribution of a response variable, which is particularly advantageous for quantifying uncertainty in predictions \cite{meinshausen2006quantile}. The QRF model was implemented through the quantile-forest Python package \cite{johnson2024quantile} with number of trees in the forest (\texttt{n\_estimators}) set to 100 and the function to measure the quality of a split as \texttt{"squared\_error"}. We computed prediction intervals by estimating quantiles ranging from 0.05 to 0.95, with the standard deviation approximated as half the difference between the upper (0.95) and lower (0.05) quantiles.

\textbf{Multilayer Perceptrons with Dropout (MLP-D)}: MLP-D is a feedforward neural network architecture incorporating Monte Carlo (MC) Dropout for uncertainty estimation \cite{gal2016dropout}. MC Dropout involves applying dropout during both training and inference, enabling the model to generate a distribution of predictions for each input, which facilitates the quantification of predictive uncertainty. In our implementation, the model architecture consisted of two hidden layers with 512 and 128 neurons, respectively, each followed by an MC Dropout layer with a dropout rate of 0.2. The ReLU activation function was used in the hidden layers to introduce non-linearity, while the output layer consisted of a single neuron for regression tasks. During training, the model was optimized using the Adam optimizer with a learning rate of 0.001 and the mean squared error loss function. The model was trained for 100 epochs, with a batch size of 32 and 10\% randomly sampled training data reserved for validation. For uncertainty estimation, we conducted 100 independent stochastic forward passes during inference, with dropout active. The mean of these predictions was used as the final output, while the 5th and 95th percentiles were computed to represent the uncertainty interval. The standard deviation of the predictions was estimated as half the difference between the 95th and 5th percentiles. The MLP-D model was implemented in TensorFlow \cite{abadi2016tensorflow} using the Keras API \cite{chollet2015keras}.

\textbf{Graph Neural Networks (GNNs)}: We implemented vanilla GNN architectures for molecular property prediction using the torch-molecule. It is a package we are actively developing to facilitate molecular discovery through deep learning approaches, featuring a user-friendly, \textit{sklearn}-style interface. This package currently supports a few prediction models, and we plan to incorporate more generative models in the near future. We implement two variants including Graph Isomorphism Networks (GIN) \cite{xu2018powerful} and Graph Convolutional Networks (GCN) \cite{kipf2016semi}. These models were trained with a batch size of 512 for 500 epochs. Hyperparameter optimization was systematically carried out through Optuna \cite{akiba2019optuna}. Key hyperparameters include the type of GNN (GIN or GCN), the normalization layer (batch, layer, or size normalization), the number of GNN layers (ranging from 2 to 5), the embedding dimension (between 256 and 512), the learning rate (from $1 \times 10^{-4}$ to $1 \times 10^{-2}$), and the dropout ratio (from 0.05 to 0.5).

\textbf{Graph Rationalization with Environment-based Augmentations (GREA)}: We employed the torch-molecule library to implement the GREA model \cite{liu2022graph}, which was trained with a batch size of 512 for 500 epochs. Hyperparameter optimization was systematically conducted using Optuna \cite{akiba2019optuna}. Key hyperparameters include the type of base encoders (GIN or GCN) for both the rationale encoder and graph encoder, the normalization layer (batch, layer, or size normalization), the number of GNN layers for the graph encoder (ranging from 2 to 5; Note: the number of layers in the rationale encoder was fixed as 2),  the embedding dimension (between 256 and 512), the learning rate (from $1 \times 10^{-4}$ to $1 \times 10^{-2}$), the dropout ratio (from 0.05 to 0.5), and the rationale subgraph size control parameter $\gamma$ (from 0.25 to 0.75). The variance of each prediction is calculated from different rationale-environment combinations in a batch and the predicted uncertainty is derived by scaling the square root of the variance by a confidence multiplier (1.96 for a 95\% confidence level).

\textbf{GPT-4o-mini for ICL Inference}: To explore the capability of LLMs in polymer property prediction, we employed GPT-4o-mini~\cite{hurst2024gpt} in zero-shot and few-shot ICL settings. Unlike traditional ML models, which require explicit training on structured datasets, LLMs leverage pre-trained knowledge and adapt to the task via contextual examples provided at inference time. In this work, GPT-4o-mini was prompted to predict polymer properties based on the SMILES representation of each test polymer. The model was evaluated under four different ICL conditions:
\begin{itemize}
    \item Zero-shot (0-ICL): No examples provided, relying entirely on the LLM's prior knowledge.
    \item Few-shot (5-ICL, 10-ICL, 20-ICL): Randomly sampled training examples (5, 10, or 20 per test instance) were included in the prompt to guide the model.
\end{itemize}

The prompts were structured to ensure a standardized response format, instructing the LLM to output only the predicted numerical value. For few-shot settings, example SMILES-property pairs were dynamically sampled from the training data for each test instance. The LLM predictions were extracted, parsed, and evaluated for accuracy.

\subsection{Interpretability}
\label{sec:method_interpretability}
To elucidate the decision-making processes of our predictive models, we employed distinct interpretability techniques tailored to both non-graph-based and graph-based architectures.

\textbf{Non-Graph-Based Models}: For MLP-D and QRF, we utilized SHAP \cite{scott2017unified} to interpret model predictions. SHAP assigns each feature an importance value for a particular prediction, grounded in cooperative game theory principles \cite{shapley1953value}. Specifically, we applied the \texttt{KernelExplainer}, a model-agnostic approach suitable for any predictive model \cite{scott2017unified}. This explainer approximates SHAP values by treating the model as a black box and perturbing input features to observe changes in the output. To enhance computational efficiency, we employed a K-means clustering algorithm on the training data to select 10 representative samples as the background dataset. The SHAP values derived from this method offer insights into contributions from those interpretable fingerprints, thereby enhancing the transparency and trustworthiness of our non-graph-based models from both local and global point of view.

\textbf{Graph-Based Model}: For GREA, we utilized employed the model-intrinsic rationalization technique developed specifically for this framework to interpret predictions \cite{liu2022graph}. In graph-based learning, a rationale refers to a subgraph that significantly influences the model's output. The GREA framework efficiently identifies such rationales by performing rationale-environment separation and enhanced representation learning in latent spaces. This method not only improves the overall learning of representations but also sharpens the identification of rationales. We highlighted the nodes with rationale score (node importance) ranking in the top 20\% of the graph as the rationale, designating these as key influencers for each polymer's property predictions.

% \subsection{Synthesizability}

% \textbf{Retrosynthesis Templates}

% \textbf{PolyScore Calculation}

% Based on the retrosynthesis process, we propose a novel synthesizability score specific to polymers, named PolyScore. Unlike existing retrosynthesis scores designed for small molecules, PolyScore is tailored to capture the unique characteristics of polymerization reactions, which considers the difficulty of synthesizing a polymer by evaluating the complexity of its retro-synthesized monomers and the number of viable polymerization routes. The PolyScore is defined as:

% \begin{equation}
% S_{\text{Poly}} = \min\left(S_{\text{Poly}}^j\right), \quad S_{\text{Poly}}^j = \text{Mean}_{\text{Reactant}} \left( S_{\text{Reactant}}^{j} \right) = \frac{n}{\sum_{i=1}^{n} \frac{1}{S_{\text{Reactant}, i}^j}} 
% \label{eq:polyscore}
% \end{equation}

% where $S_{\text{Reactant}, i}^j$ represents the synthetic accessibility score (SAScore) of the $i^{th}$ retro-synthesized monomer in the $j^{th}$ route. The final PolyScore ($S_{\text{Poly}}$) is taken as the minimum of the scores across all possible polymerization routes ($S_{\text{Poly}}^j$). This approach ensures that polymers with easier synthesis routes receive a higher score, reflecting their higher likelihood of successful synthesis in practice.

\section{Data and code availability}
All data and code used in this study are available at \href{https://github.com/Jiaxin-Xu/POINT2.git}{https://github.com/Jiaxin-Xu/POINT2.git}.
% \section{Author contributions}
\section{Conflicts of interest}
The authors declare no conflicts of interest.
\section*{Acknowledgments}
This work was supported in part by National Science Foundation (2102592, 2332270). 

%Bibliography
\bibliographystyle{unsrt}  
\bibliography{main}  
% \newpage

\appendix
\renewcommand{\thetable}{\Alph{section}.\arabic{table}}
\renewcommand{\thefigure}{\Alph{section}.\arabic{figure}}

\section{Additional Results}

\begin{figure}[h]
    \centering
    \includegraphics[width=1\linewidth]{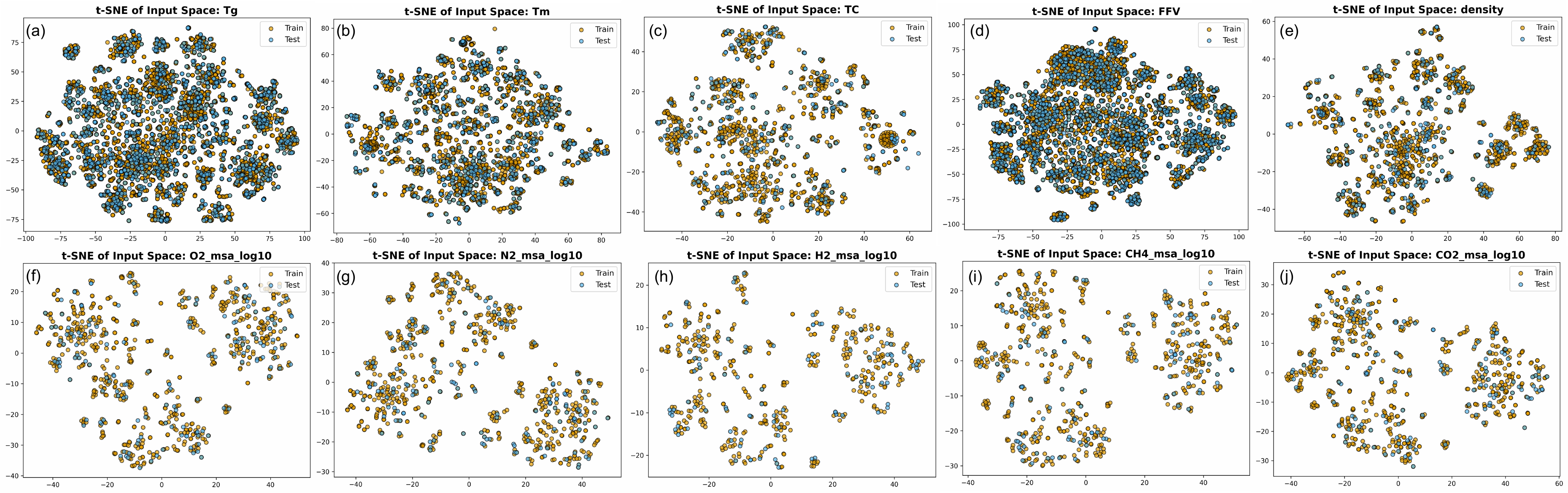}
    % \vspace{-0.1in}
    \caption{T-SNE plot of the input space comparison between training and testing data for properties: (a) T\textsubscript{g}, (b) T\textsubscript{m}, (c) TC, (d) FFV, (e) $\rho$, (f) P(O$_2$), (g) P(N$_2$), (h) P(H$_2$), (i) P(CH$_4$), and (j) P(CO$_2$).}
    \label{fig:tsne}
    % \vspace{-0.25in}
\end{figure}

\begin{figure}[h]
    \centering
    \includegraphics[width=1\linewidth]{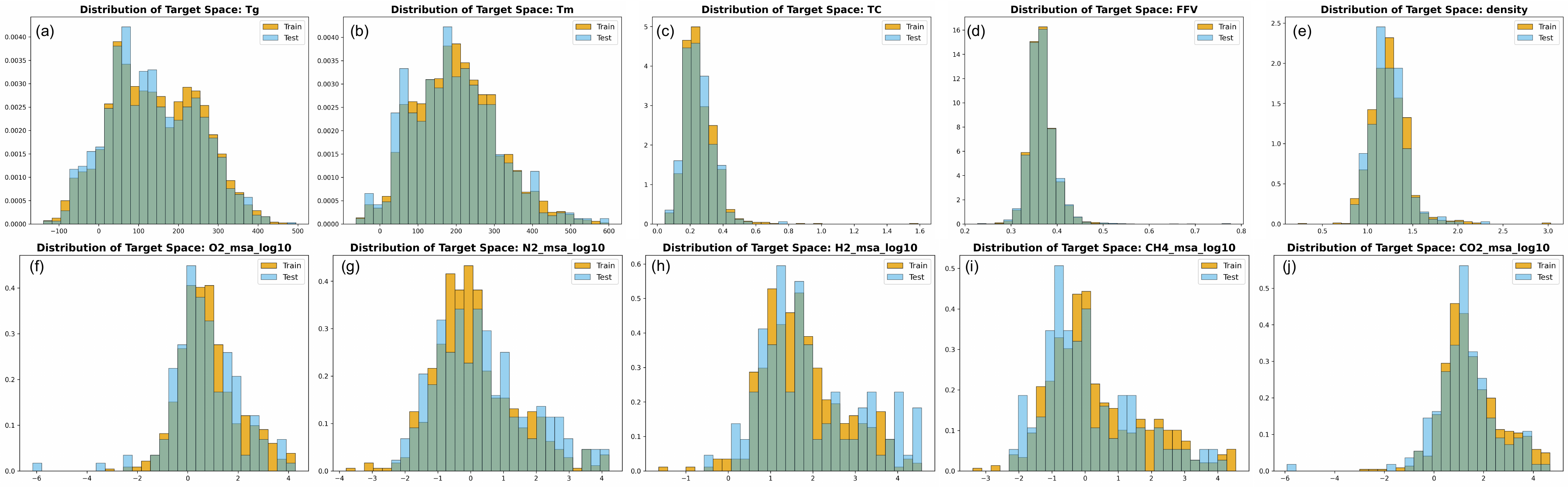}
    % \vspace{-0.1in}
    \caption{Label distribution of training and testing data for properties: (a) T\textsubscript{g}, (b) T\textsubscript{m}, (c) TC, (d) FFV, (e) $\rho$, (f) P(O$_2$), (g) P(N$_2$), (h) P(H$_2$), (i) P(CH$_4$), and (j) P(CO$_2$).}
    \label{fig:label_dist}
    % \vspace{-0.25in}
\end{figure}

\begin{table}[ht]
\centering
\caption{Summary of average RMSE values for different polymer fingerprints (across QRF and MLP-D) and graph representation (across GNN and GREA) on the test datasets. Best fingerprint (except graph) for each task is in bold.}
\label{tab:fingerprint_averages}
\begin{tabular}{lcccccccccc}
\toprule
\textbf{Fingerprint} & T$_g$ & T$_m$ & TC & FFV & $\rho$ & P(O$_2$) & P(N$_2$) & P(H$_2$) & P(CH$_4$) & P(CO$_2$) \\
\midrule
Morgan  & \textbf{37.82} & 59.63 & 0.0515 & 0.0155 & 0.1120 & \textbf{0.6690} & 0.5335 & 0.4570 & 0.5980 & \textbf{0.6680} \\
MACCS   & 41.98 & 61.11 & 0.0570 & \textbf{0.0145} & \textbf{0.1025} & 0.7275 & 0.6140 & \textbf{0.4000} & 0.7555 & 0.7135 \\
RDKit   & 38.71 & \textbf{59.50} & 0.0565 & 0.0150 & 0.1220 & 0.7275 & 0.5390 & 0.4890 & 0.5995 & 0.7780 \\
TT      & 39.32 & 62.82 & \textbf{0.0490} & 0.0165 & 0.1540 & 0.7470 & 0.5300 & 0.5015 & \textbf{0.5725} & 0.8015 \\
AP      & 39.30 & 60.55 & 0.0505 & 0.0150 & 0.1155 & 0.6785 & \textbf{0.5145} & 0.4755 & 0.6005 & 0.7010 \\
\midrule
\rowcolor{gray!20}
Graph & 36.67 & 56.29 & 0.0715 & 0.0220 & 0.1470 & 0.5870 & 0.4995 & 0.4580 & 0.5085 & 0.626 \\
\bottomrule
\end{tabular}
\end{table}

\begin{figure}[h]
    \centering
    \includegraphics[width=1\linewidth]{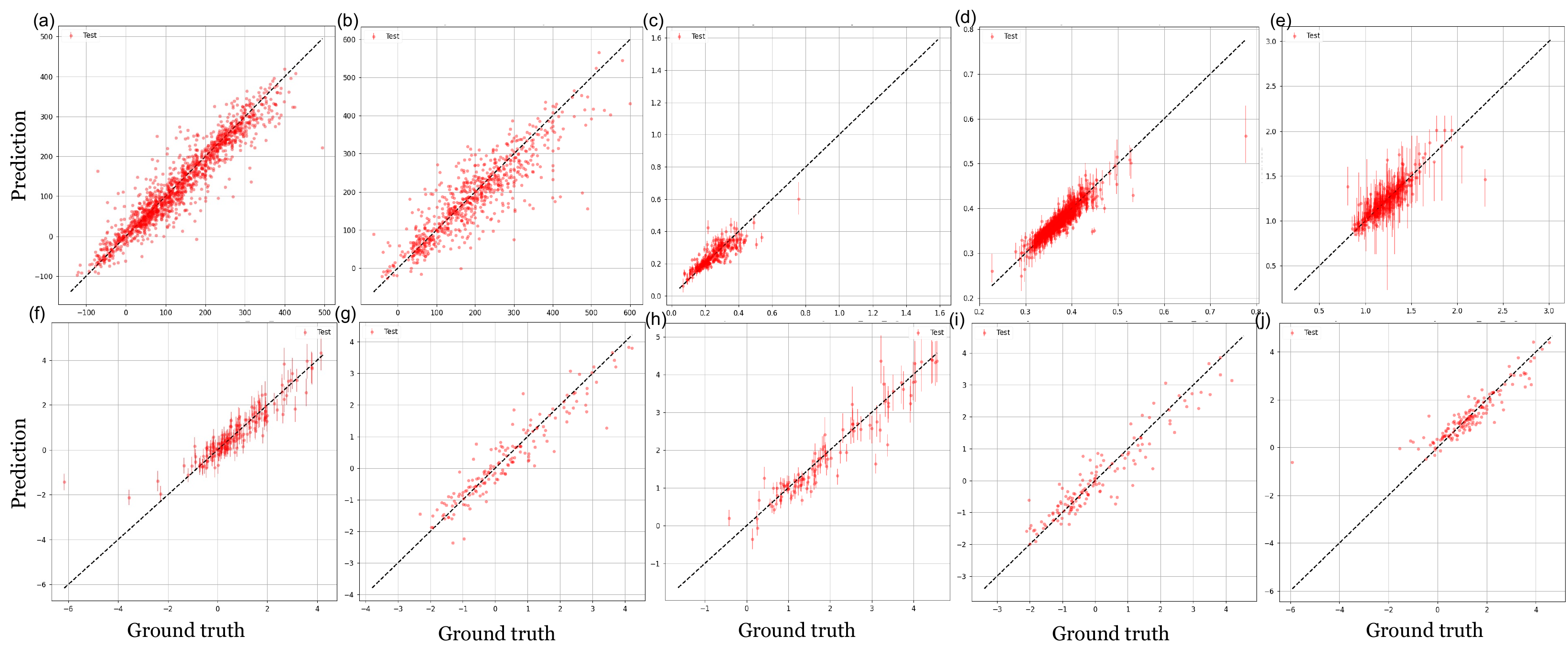}
    % \vspace{-0.1in}
    \caption{Prediction parity plots of the best model on test dataset of properties: (a) T\textsubscript{g} (GNN), (b) T\textsubscript{m} (GNN), (c) TC (MLP-D-Morgan), (d) FFV (MLP-D-MACCS), (e) $\rho$ (QRF-MACCS), (f) P(O$_2$) (GREA), (g) P(N$_2$) (GNN), (h) P(H$_2$) (MLP-D-MACCS), (i) P(CH$_4$) (GNN), and (j) P(CO$_2$) (GNN). X-axis represents the ground truth and y-axis represents the prediction. All values are in units of the corresponding label unit in Table \ref{tab:dataset}.}
    \label{fig:test_parity}
    % \vspace{-0.25in}
\end{figure}

\begin{figure}[h]
    \centering
    \includegraphics[width=1\linewidth]{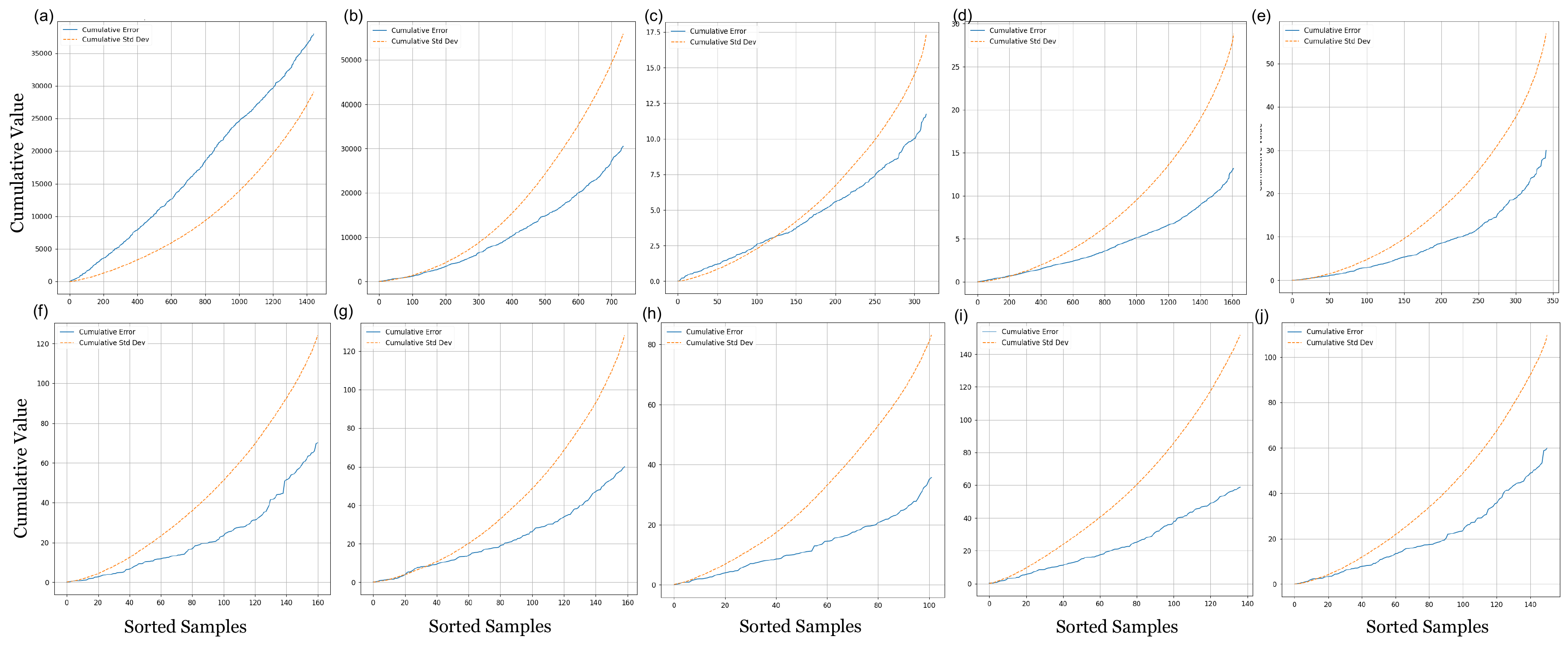}
    % \vspace{-0.1in}
    \caption{Sparsification plots of the best model on test dataset (in terms of $\rho_s$) of properties: (a) T\textsubscript{g} (MLP-D-TT), (b) T\textsubscript{m} (QRF-TT), (c) TC (QRF-Morgan), (d) FFV (QRF-TT), (e) $\rho$ (QRF-TT), (f) P(O$_2$) (QRF-TT), (g) P(N$_2$) (QRF-TT), (h) P(H$_2$) (QRF-AP), (i) P(CH$_4$) (QRF-AP), and (j) P(CO$_2$) (QRF-Morgan). X-axis represents test data sample IDs, ranked in descending order of predicted uncertainty (i.e., from lowest to highest predicted uncertainty) and y-axis represents the the cumulative value of prediction error (blue solid line) and prediction uncertainty (standard deviation, orange dashed line), which is in units of the corresponding label unit in Table \ref{tab:dataset}.}
    \label{fig:test_sparsity}
    % \vspace{-0.25in}
\end{figure}

\end{document}